\documentclass[%
superscriptaddress,
bibnotes,
 amsmath,amssymb,
 aps,
 prb, 
longbibliography,
eprint,
floatfix,
reprint
]{revtex4-2}

\usepackage{xcolor}
\definecolor{darkblue}{RGB}{0,0,150}
\definecolor{nightblue}{RGB}{0,0,100}

\usepackage{mathrsfs,dsfont,mathtools,braket}

\newcommand{\refsub}[2]{\hyperref[#1]{\ref*{#1}#2}}

\usepackage{bm}
\usepackage[
colorlinks,
citecolor=darkblue,
linkcolor=darkblue,
urlcolor=nightblue]{hyperref}

\usepackage[english]{babel}
\usepackage[babel,kerning=true,spacing=true]{microtype}

\definecolor{DarkRed}{RGB}{100,0,0}
\definecolor{LightGreen}{RGB}{000,50,0}

\newcommand{\f}{\frac}
\def\({\left(}
\def\){\right)}
\def\prl{\partial}
 \newcommand{\D}{{\rm d}}

\newcommand{\lb}{\left[}
\newcommand{\rb}{\right]}
\newcommand{\half}{\frac{1}{2}}

\def\inf{\infty}

\newcommand{\beq}{\begin{equation}}
\newcommand{\eeq}{\end{equation}}
\newcommand{\bal}{\begin{aligned}}
\newcommand{\eal}{\end{aligned}}

\newcommand{\bfa}{{\bf a}}
\newcommand{\bfb}{{\bf b}}
\newcommand{\bfk}{{\bf k}}
\newcommand{\bfq}{{\bf q}}
\def\a{\alpha}
\def\d{\delta}
\def\De{\Delta}
\def\e{\epsilon} 
\def\E{\hbox{$\cal E $}}
\def\s{\sigma}
\def\r{\rho}
\newcommand{\vp}{\varphi}
\def\w{\omega}

\begin{document}
\title{
The quantum geometric origin of capacitance in insulators
}
\author{Ilia Komissarov}
\email{i.komissarov@columbia.edu}
\affiliation{Department of Physics, Columbia University, New York, NY 10027, USA}
\author{Tobias Holder}
\email{tobiasholder@tauex.tau.ac.il}
\affiliation{School of Physics and Astronomy, Tel Aviv University, Tel Aviv, Israel}
\affiliation{Department of Condensed Matter Physics, Weizmann Institute of Science, Rehovot, Israel}
\author{Raquel Queiroz}
\email{raquel.queiroz@columbia.edu}
\affiliation{Department of Physics, Columbia University, New York, NY 10027, USA}
\affiliation{Center for Computational Quantum Physics, Flatiron Institute, New York, New York 10010, USA}

\date{\today}

\begin{abstract}
In band insulators, without a Fermi surface, adiabatic transport can exist due to the geometry of the ground state wavefunction.
Here we show that for systems driven at a small but finite frequency $\omega$, transport likewise depends sensitively on quantum geometry. We make this statement precise by expressing the Kubo formula for conductivity as the variation of the time-dependent polarization with respect to the applied field. We find that at linear order in frequency, the longitudinal conductivity results from an intrinsic capacitance determined by the ratio of the quantum metric and the spectral gap, establishing a fundamental link between the dielectric response and the quantum metric of insulators. We demonstrate that quantum geometry is responsible for the electronic contribution to the dielectric constant in a wide range of insulators, including the free electron gas in a quantizing magnetic field, for which we show the capacitance is quantized. We also study gapped bands of hBN-aligned twisted bilayer graphene and obstructed atomic insulators such as diamond. In the latter, we find its abnormally large refractive index to have a topological origin.
\end{abstract}

\maketitle

\section*{Introduction}

Historically, the primary focus when examining material properties has been the electronic band structure. However, following the transformative impact of the modern theory of polarization \cite{Vanderbiltbook}, the Hilbert space geometry has emerged in recent years as a critical instrument for characterizing quantum materials.
This perspective shift has been fueled mainly by the rapid progress in the understanding of the quantum geometric tensor (QGT)~\cite{Provost1980}, whose imaginary part, the Berry curvature, has become indispensable in addressing topological band structure properties~\cite{Xiao2010,Hasan2010, Qi2011}.  
On the other hand, the real part, known as the quantum metric, has only recently attracted attention. This quantity was shown to appear in various transport functions including optical conductivity~\cite{Gao2015,Kolodrubetz2017,Gao2019a,Holder2020,Mitscherling2020,Ahn2022,Bouzerar2022,Mitscherling2022,Chen2022,deSousa2023,Onishi_2024}, and may contribute to the superfluid stiffness in flatband superconductors~\cite{Peotta2015, Liang2017, Hu2019, Julku2020}. 

Here, we add to this growing list a seemingly overlooked property of the QGT, which is how it enters canonically in the quasistatic conductivity in insulators. Since the real part of QGT indicates the spread of the Wannier functions, it is expected to enter the dielectric properties of the materials such as the response to the polarizing electric field. This perspective is motivated by the modern theory of polarization:
The polarization $P^\mu$ can be defined by carefully evaluating the position operator in momentum space~\cite{Souza2000,Souza2004,Resta2011}. This implies that $P^\mu$ is determined entirely by the eigenfunctions of a given Bloch-periodic Hamiltonian and independent of the eigenvalues (dispersion).
Since the current can be defined as a derivative of the polarization $j^\mu=dP^\mu/dt$, it is tempting to seek similar conclusions for $j^\mu$. 
This expectation is indeed true for a band insulator, where the dc-current is purely transverse, dissipationless and proportional to the Chern number of the ground state. 

Here, we investigate how the linear response in an insulator 
changes away from the zero-frequency limit, establishing that the quasistatic expansion for low frequencies contains valuable additional information about the quantum geometry of the Hilbert space. This insight allows us to connect the longitudinal conductivity with the 
quantum metric in insulators where the typical bandwidth is small compared to the band gap.

\begin{figure}
\centering
\includegraphics[width=\columnwidth]{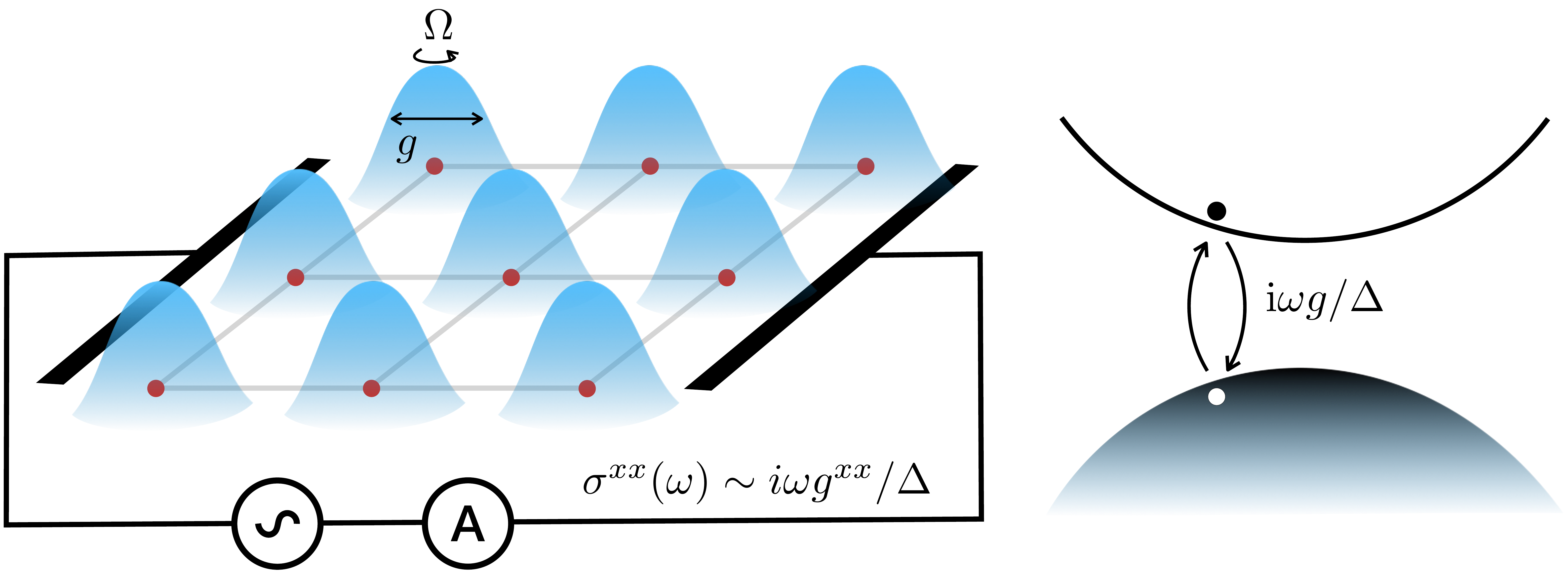}
\caption{Emergence of the intrinsic capacitance in insulators with nonzero quantum metric $g$. Upon applying an ac electric field, due to virtual excitations a transient polarization is induced, which entails a longitudinal, purely imaginary impedance.}
\label{fig:fig1}
\end{figure}

The starting point for our considerations is the static susceptibility $\chi$ in insulators expressed in terms of the polarization as
$\chi^{\mu\nu}=\partial P^\nu/\partial E^\mu$~\cite{Vanderbiltbook}.
Using the definition of the current, one can similarly evaluate the conductivity as $\sigma^{\mu\nu}\to\partial j^\nu/\partial E^\mu$.
However, since the current arises in a quasisteady state, it is not obvious to which extent the properties of a static polarization carry over to the conductivity~\cite{Nenciu1991,Souza2004,Marzari2012}.
We show by explicit construction using the Kubo formula that linear conductivity $\sigma$ in both insulators and metals can be expressed in terms of the polarization as 
\begin{align}
\sigma^{\mu\nu}(\omega) 
&= 
\frac{\delta}{\delta E^\mu} 
\biggl(\frac{d P^\nu}{d t}
\biggr)\bigg|_{E^\mu=0} \, ,
    \label{eq:mainKubo}
\end{align}
where the time dependence of the polarization enters through the monochromatic electric field $E_\mu(t)=E_\mu e^{i\omega t}$. 
Eq.~\eqref{eq:mainKubo} makes use of a functional derivative with respect to the time-dependent field. 
Based on this insight, we are motivated to explore the geometric content of the quasistatic response. 
 Focusing on two-dimensional insulators with rotational symmetry, the low-frequency expansion yields
\begin{align}
    \sigma^{\mu\nu}(\omega)
    &=-\frac{e^2}{h}C\epsilon^{\mu\nu}
    +i\omega c \, \delta^{\mu \nu}+\dots\, ,
    \label{eq:maincapacitance}
\end{align}
with $c$ being related to the static susceptibility in three dimensions via the vacuum permittivity $\epsilon_0$ as $c = \epsilon_0 \chi$, and $C$ denoting the Chern number.
Note that $c$ is related to the capacitance of a three-dimensional parallel-plate capacitor with cross-sectional area $A$ and distance between the plates $d$ by $Ac/d$.
That is, the capacitance $c$ constitutes the leading low-frequency longitudinal contribution, indicative of the deviation from the static response.
This raises two important questions: Does this quantity depend on the quantum geometry? And is $c$ a substantial contribution in quantum materials?
This letter answers both questions with yes.
In particular, we present several examples of non-interacting band insulators where the intrinsic capacitance $c$ contains the matrix elements of the quantum metric, normalized by the respective band gaps.
The appearance of the quantum metric in the quasistatic conductivity can be understood intuitively by recalling that the quantum metric quantifies the extent of a wavefunction in real space~\cite{Marzari1997,  Marzari2012}. Therefore, it measures how much the electrons can polarize. In short, systems with a finite quantum metric exhibit an intrinsic capacitance.
This capacitance is a purely quantum phenomenon that arises from coherent, virtual interband transitions between the full valence band and the empty conduction band~\cite{Holder2021}, simply because within a full band quasiparticles cannot be displaced at all (cf. Fig.~\ref{fig:fig1}). 

We emphasize that the intrinsic capacitance $c$ originates solely from the multi-banded nature of insulators, making it the only electronic contribution to the geometric capacitance of clean insulators at frequencies smaller than the band gap. In dirty or doped insulators, due to in-gap bound states \cite{Viehweger1991} the measured capacitance is augmented by the quantum capacitance, which is proportional to the density of states at the Fermi level inside the mobility gap.
As the trace of the quantum metric is bounded by the Chern number~\cite{Roy2014a, Peotta2015, Ozawa2021}, we expect the capacitive response to be enhanced in systems where the orbitals cannot be exponentially localized. 
In the following, we clarify under which circumstances the intrinsic capacitance can be used as a diagnostics of the quantum geometry of the system. 

\section*{Results}

\emph{Geometric expansion of the Kubo formula. ---}
It is well established that in the adiabatic transport regime, the response of an insulator to external fields is dictated by the geometry of the Hilbert space. Eq.~\eqref{eq:mainKubo} indicates that beyond the adiabatic regime, we may view the conductivity solely as a geometric quantity in the Hilbert space slowly evolving in time. This statement is now made precise. To this end, let us consider a periodic system with Bloch wavefunctions $|m \bm{k}\rangle$ and energy eigenvalues $\hbar\omega_{m}(\bm{k})$, where $m$ is the band index, and in the following we will suppress the momentum index. Using the Kubo formula, we find that any response at finite frequency depends explicitly on the band dispersion of filled and empty bands. This dependence is introduced by the interband elements of the current operator $J_{nm}^\mu$ with spatial index $\mu$, which are explicitly given by $J_{nm}^\mu=-ie(\omega_n-\omega_m)r^\mu_{nm}\equiv -ie\omega_{nm}r^\mu_{nm}$, with $r^\mu_{nm}=\langle n|\hat r^\mu|m\rangle$ representing the matrix elements of the position operator. In the insulating state, the conductivity exclusively depends on interband terms which are given by
\beq
\sigma^{\mu \nu} (\omega)= - \frac{i e^2}{\hbar} \sum\limits_{n \neq m} \int_{\rm BZ} f_{nm} \omega_{nm}  \frac{  r^\mu_{nm}  r^\nu_{mn}}{ \omega_{nm}+\omega} \, ,
\label{k31t}
\eeq
where we introduced the difference of occupation functions $f_{nm}=f_n-f_m$. The integral is over the Brillouin zone, and all band structure quantities implicitly depend on the momentum unless specified otherwise.
As we further detail in the supplementary information~\cite{SM}, this expression can be exactly rewritten as 
\beq
\sigma^{\mu \nu} (\omega)=  -\frac{i e^2}{\hbar} \int_{\rm BZ} \int_{\rm C} e^{- i \w_+ t} \frac{d}{d t} \left( \hat T Q^{\mu \nu}(t) \right)\label{KuboTimeQGT}
\eeq
where we introduced a time-dependent QGT
\beq
Q^{\mu \nu}(t) = \sum\limits_{n \neq m} f_n(1-f_m) r^\mu_{nm}(0)  r^\nu_{mn}(t)\, .
\eeq
Note that the integral in \eqref{KuboTimeQGT} is evaluated over the Keldysh contour $\rm C$~\cite{SM}, and $\hat T$ denotes path ordering. 
$Q(t)$ can be written succinctly in operator form as $Q^{\mu \nu}(t) \equiv \mathrm{Tr}[\hat  P \hat r^\mu(0) (1-\hat P) \hat r^\nu (t) ]$~\cite{HerzogArbeitman2022}, where $\hat P$ is the projector into the filled bands. In this form, the time-dependent QGT can be clearly identified as a generalization of the time-independent QGT.

The intrinsic capacitance arises in the Kubo formula upon expansion to linear order in the driving frequency. We assume that the frequency is well below the band gap, and therefore transport is non-dissipative. Let us concentrate on the longitudinal conductivity that, in the absence of dissipation, is purely imaginary. Expanding in powers of frequency $\sigma^{\mu\mu} \simeq i \omega c^{\mu\mu} + O(\omega^3)$, we extract the capacitance
\begin{align}
\label{CapWneq0}
c^{\mu\mu} = \frac{2 e^2}{\hbar}  
\int_{\rm BZ} \sum_{m\neq n}
f_n(1-f_m) \frac{ g^{\mu\mu}_{mn}}{\omega_{mn}} \, ,
\end{align}
where the numerator contains the matrix elements of the quantum metric
$g^{\mu \mu}_{mn} = \{r^\mu_{nm},r^\mu_{mn}\}/2$, from which the full ground state quantum metric can be obtained by the summation $g^{\mu\mu}=\sum_{nm}f_n(1-f_m) g_{mn}^{\mu\mu}$. For isotropic systems, we drop the spatial indices, such that $c\equiv c^{\mu\mu}$. 
The identification of an intrinsic capacitance due to the quantum metric, Eq.~\eqref{CapWneq0}, is the main result of this work.

\emph{Landau Levels. ---}
We first consider the intrinsic capacitance in the well-known case of an electron gas under an applied out of plane magnetic field $B$. The spectrum of this problem consists of flat Landau levels with a uniform gap given by the cyclotron frequency $\omega_{mn}= \omega_c =e B/m_e$. Since the dipole transitions are only allowed between neighboring Landau levels, i.e. $r^\mu_{nm}\propto\delta_{n+1,m}$, Eq.~\eqref{CapWneq0} simplifies dramatically. The quantum metric for the case of Landau levels is given by $g^{xx} =
l_{\rm B}^2 C/2$ \cite{Ozawa2021} with $l_{\rm B} = \sqrt{\hbar/e B}$, such that the capacitance takes the quantized value of~\cite{SM}
\beq
c = \frac{e^2}{h \omega_c} C \, . 
\label{LanCap}
\eeq
Most importantly, this implies that at sufficiently small frequency, each filled Landau level carries a quantum of capacitance $c_0 = e^2/h \omega_c$. The reason for that can be gleaned from Eq.~\eqref{CapWneq0}: in a system with flatband dispersion, the energy difference $\omega_{mn}$ can be taken out of the momentum integral, such that only the integral over $g_{mn}$ remains. For a magnetic field of $1\,$T, we find $c_0 \simeq 0.22\, \mathrm{fF}$, which is well within the range of Microwave Impedance Microscopy (MIM) devices \cite{Wu2010,Cui2016}.

\begin{figure}
\includegraphics[width=.45\textwidth]{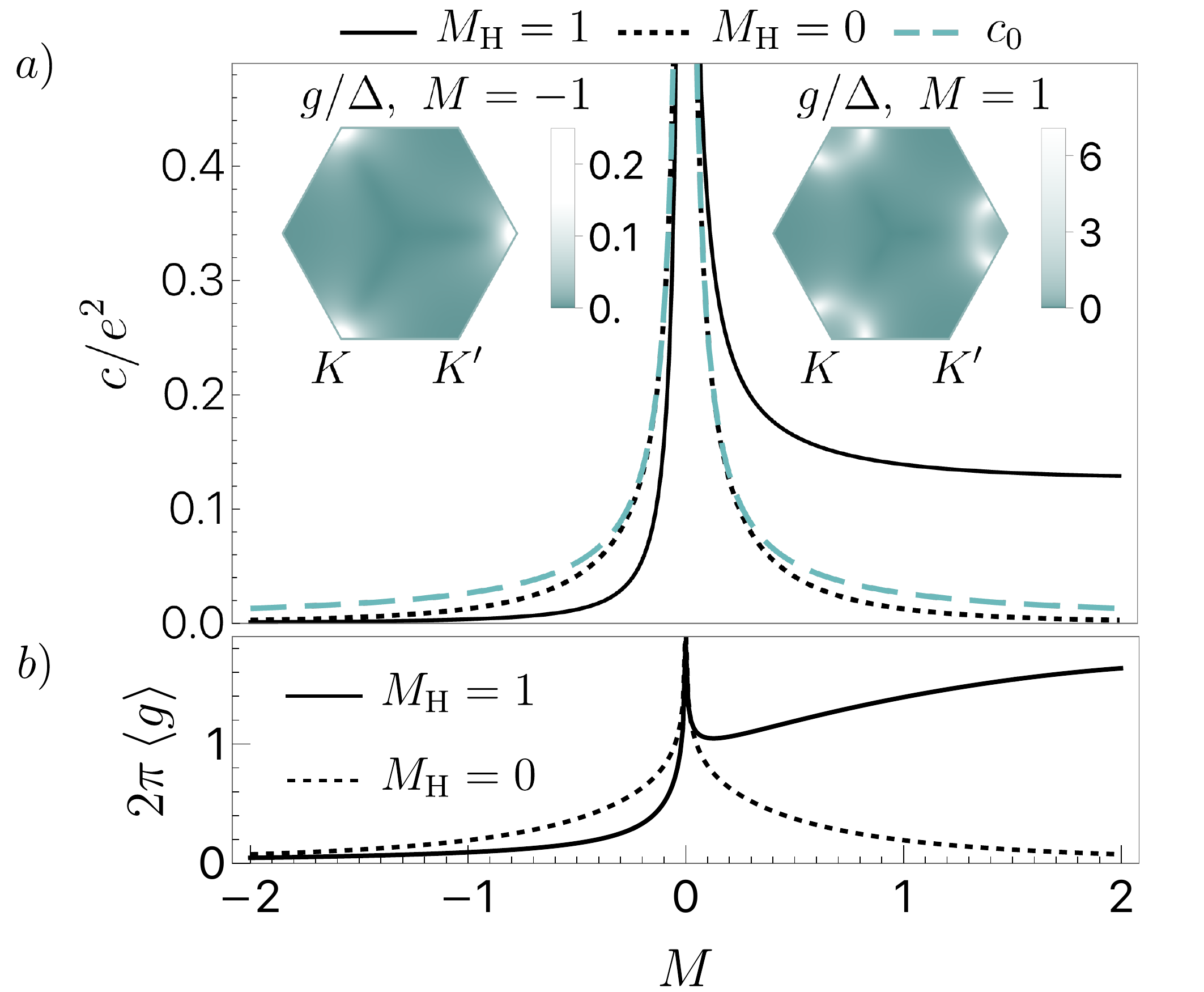}
     \caption{Relation between the quantum metric and intrinsic capacitance $c$ in a Haldane model (see Ref.\cite{SM}) compared against the linear Dirac cone result \eqref{cDirac}. $a)$ The top panel displays $c$ as a function of the mass parameter $M$. The solid curve corresponds to the value of the Haldane mass $M_{\rm H} = 1$, while the black dashed curve represents $M_{\rm H} = 0$. The colored dashed curve represents the capacitance $c_0$ of a linear Dirac cone. The insets on the left and right show the relative contributions to $c$ across the Brillouin zone in the Haldane $M_{\rm H}=1$ model with $M = -1$ (trivial) and $M = 1$ (topological), respectively. The involvement of the high-momentum modes for $M=1$ confirms the topological nature of transport in this parameter region.
    $b)$ Integrated, normalized, traced over the spatial indices quantum metric $\braket{g}=
    \int_{\rm BZ} (g^{xx}+g^{yy})$ in the Haldane model with $M_{\rm H}=1$ (solid line), and for $M_{\rm H}=0$ (dashed line). Both $c$ and $\left< g\right>$ for $M_{\rm H}=0$ are divided by a factor of two for better comparison, since in this case at $M=0$ the gap closes in both valleys.}
     \label{fig:haldane}
\end{figure}

\emph{Gapped Dirac Hamiltonians. ---} 
We consider a two-dimensional gapped Dirac dispersion, given by the continuum Hamiltonian $\hat H = k^\mu  \hat \sigma_\mu + M \hat \s_z$.
The longitudinal ac conductivity~\eqref{k31t} simplifies to
\beq
\label{sxxIntDirac}
\s(\omega) = {i e^2 \omega} \int_{\rm BZ} \frac{\Delta g}{\Delta^2 - (\hbar \omega)^2}\, ,
\eeq
with the gap given by $\De = 2 \sqrt{M^2 + k^2}$~\cite{SM}, and $ g = g^{xx}+g^{yy}$. 
For low frequencies, far smaller than the gap, the linear coefficient of $i\omega$ corresponds to the capacitance of a massive Dirac fermion 
\beq
\label{cDirac}
c_0 = \frac{e^2}{12 \pi |M|}\,.
\eeq
This quantity approximates well the contribution to the capacitance per Dirac cone in topologically trivial materials.
To consider the influence of topology on this value, we consider a Dirac Hamiltonian with a quadratic correction $\hat H' = k^\mu  \hat \sigma_\mu + (M - \alpha k^2) \hat \s_z$, which yields for the capacitance
\beq
\label{conek2}
c_0' = 
\begin{cases} 
        \frac{e^2}{12 \pi} \frac{1}{|M|  |1 - 4 M \alpha|}\, , & \a M < 0 \, , \\
      \frac{e^2}{12 \pi |M|} + \frac{|\alpha|}{6 \pi}\, , &  \a M > 0\, . \\
   \end{cases}
\eeq
If $M$ and $\alpha$ have the same sign, $\hat H'$ describes a Chern insulator, with the capacitance acquiring a minimum value $c_{\rm min}=|\alpha|/6\pi$ in the limit $M\to\infty$. This behavior is reminiscent of the well-known bound on the quantum metric $2 \pi \int_{\rm BZ} g \geq |C|$~\cite{Roy2014a, Peotta2015, Ozawa2021}. 
In the trivial region $\a M<0$, $c_0'$ decays faster than $c_0$. To summarize, the intrinsic capacitance diverges at a gap-closing, and acquires a characteristic asymmetry between trivial and topological phase, but only at a finite distance to the topological transition, while very close to the transition, $c_0'\simeq c_0$ is symmetric.

To illustrate how these insights carry over to the tight-binding models of topological materials, we calculate the capacitance in a Haldane-like model parameterized by the sublattice staggered potential $M$, both with and without the Haldane mass term $M_{\rm H}$~\cite{SM}. 
As shown in Fig.~\ref{fig:haldane}, for finite $M_{\rm H}$, the system experiences a transition between the trivial phase ($M<0$) and the topological phase ($M>0$). 
As the gap closes at $M=0$, $c$ diverges on both sides of the phase transition. Close to the transition, $c$ is well approximated by the Dirac cone result $Nc_0$, where $N$ is the number of gapless valleys at $M=0$: $N=2$ for $M_{\rm H}=0$, and $N=1$ for $M_{\rm H}=1$. 
Similar to what we have demonstrated for $c_0'$, the intrinsic capacitance is not symmetric across both phases: it saturates at a finite value on the topological side of the Haldane model but quickly decays to zero on the trivial side. 
In contrast, for zero Haldane mass ($M_{\rm H}=0$), $c$ remains symmetric deep into the gapped phase.
These features of the intrinsic capacitance are due to its interband origin, which makes it sensitive both to geometric and topological properties of the band structure.

\emph{Magic angle twisted bilayer graphene ---} Let us now consider the magic-angle twisted bilayer graphene aligned on top of hBN~\cite{Cea2020,Shi2021}. Due to the alignment, the in-plane inversion symmetry is broken, and the bands close to charge neutrality acquire a gap $\Delta$ of the order 10 meV \cite{Liu2022}: this configuration makes the capacitance at half filling well-defined.

We calculate the capacitance $c$ numerically in the Bistritzer-MacDonald model \cite{Bistritzer2011} with a Fermi level at charge neutrality for realistic interlayer $AA$- and $AB$-sublattice tunneling parameters $u=0.077\, \rm eV$ and $u'=0.11\, \rm eV$ \cite{senTBG}. The presence of the substrate is mimicked by a sublattice-polarized local term of strength $\De/2$. We are interested in the value of $c$ in a range of twist angles $\theta$ in the vicinity of the ``magic'' $\theta_{\rm M} \simeq 1.063^\circ$.  

\begin{figure}[t!]
\centering
\includegraphics[width=.45\textwidth]{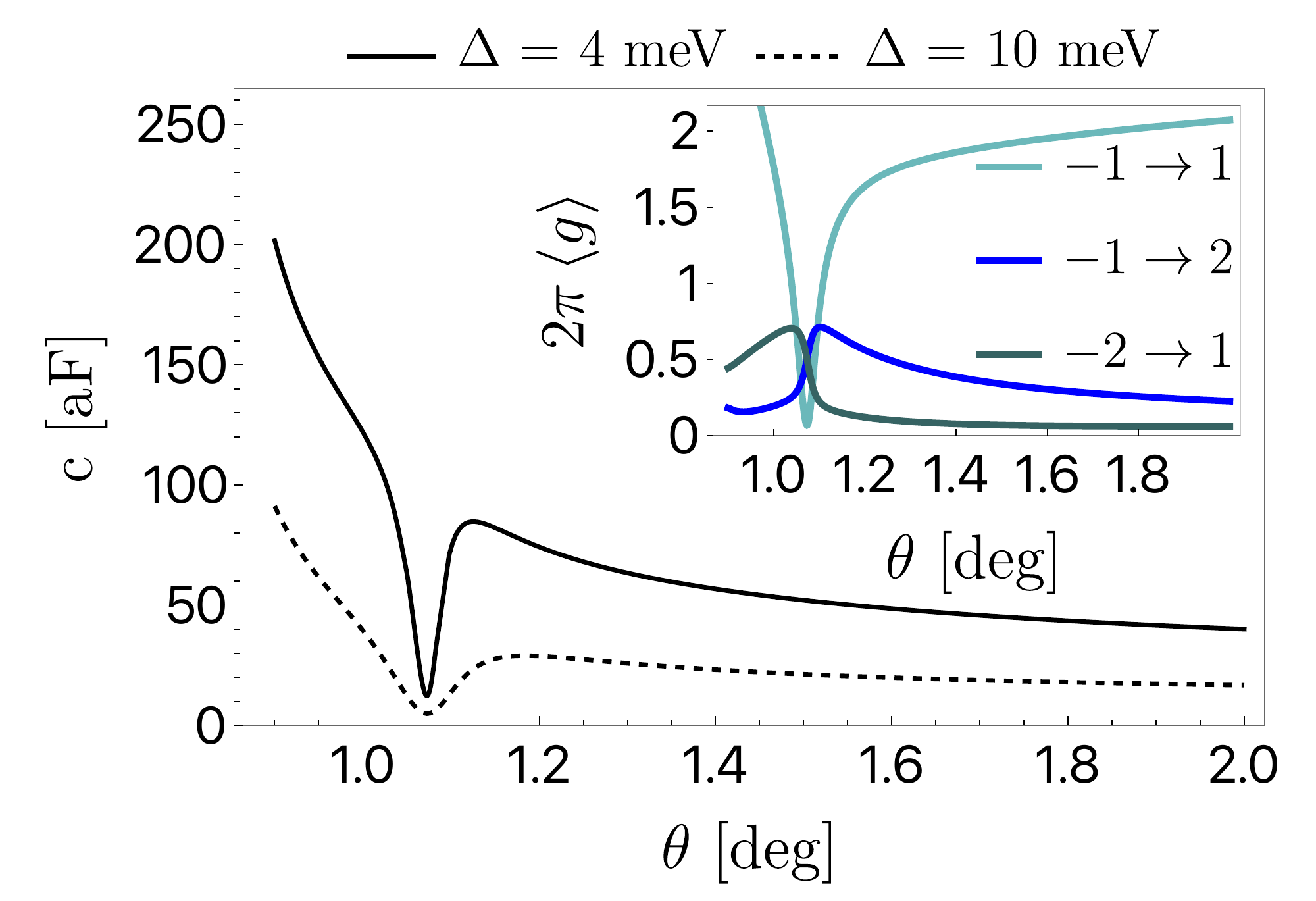}
     \caption{Intrinsic capacitance $c$ for hBN-aligned twisted bilayer graphene, depicted in relation to the twist angle $\theta$ for two values of the gap between the flat bands. In the inset, we sketch the behavior of the interband matrix elements of the quantum metric, integrated over the momentum in one valley for the case $\Delta = 4~\rm meV$. The band indices are measured from charge neutrality, with $-1$ and 1 corresponding to the flat bands.}
     \label{fig:TBGpic}
\end{figure}

We present the resulting behavior of $c(\theta)$ in Fig.~\ref{fig:TBGpic} for two values of $\De$. Since the dispersion in the flat bands is minimized at $\theta_{\rm M}$, based on Eq.~\eqref{CapWneq0} one would expect a local maximum of the capacitance to appear at the magic angle. Contrary to this expectation, this quantity develops instead a sharp local minimum since an abrupt decrease in the quantum metric of the flat bands around $\theta_{\rm M}$ dominates upon band flattening. The quench of the quantum metric due to the saturation of the trace condition is known to occur in the chiral limit of Bistritzer-MacDonald model \cite{Ledwith_2022} and expected to hold by continuity away from the $u \to 0$ limit. Interestingly, the value of the capacitance at the minimum can be roughly estimated as \eqref{cDirac} multiplied by the number of Dirac quasiparticles, which for $\De=4\,$meV gives $8 c_0 \simeq 17\, {\rm aF}$.

We see that the behavior of the intrinsic capacitance $c$ correctly captures the distinctive and sudden decrease of the wavefunction spread that takes place when the bands are flattened, and therefore, the response $c$ emerges as an efficient probe of the quantum metric. We expect the suppression of the dielectric constant with ensuing anti-screening effects to play a role in stabilizing the interacting phases observed at the magic angle \cite{pjh, Xie_2021}. We leave the investigation into the relevance of our findings to correlated phenomena for future work.

\emph{Electronic contribution to the dielectric constant ---} 
Having established the intrinsic capacitance as a valuable observable to diagnose the quantum geometry of systems with nearly flat bands, we address the question of what information $c$ contains in the case of generic insulators with a dispersive spectrum. To this end, we point out that the (dimensionless) electronic component of the dielectric constant in linear response theory is related to $c$ as $\epsilon^{\mu\nu} = \d^{\mu \nu} + c^{\mu \nu}/ \epsilon_0$, and therefore given by
\beq
\label{eq:diel}
\epsilon^{\mu\nu}\!=\!\delta^{\mu\nu}+
\frac{2 e^2}{\hbar \epsilon_0}   \sum_{m\neq n} \int_{\rm BZ}
\!\!f_n(1-f_m) \frac{ g^{\mu\nu}_{mn}}{\omega_{mn}}\, ,
\eeq
which implies in particular that the diagonal elements of the dielectric constant are bound by the integral of the metric over the Brillouin zone and the spectral gap $\Delta$
\beq
\label{eq:bd}
\epsilon^{\mu\mu}\leq 1+
\frac{2 e^2}{\epsilon_0} 
 \frac{ \int_{\rm BZ} g^{\mu\mu}}{\Delta} \, .
\eeq
The expression on the right can be determined solely from the knowledge of the ground state quantum metric and the gap, and therefore can be especially useful in strongly correlated insulators where the full excitation spectrum is unknown. As an example, in fractional Hall insulators, the optical gap takes a uniform value $\hbar \omega_c$ due to Kohn's theorem \cite{KohnT}, such that the bound \eqref{eq:bd} is saturated, and we obtain
\begin{equation}
\label{eq:fqh}
(\epsilon^{\mu\mu})_{\rm FQHE}= 1+
\frac{2 e^2}{\epsilon_0} 
 \frac{\int_{\rm BZ} ( g^{\mu\mu} )_{\rm FQHE}}{\hbar \omega_c}\, .
\end{equation}

\emph{Estimating the quantum metric from $\epsilon$ ---}
Motivated by the relation \eqref{eq:diel}, one may ask whether the localization properties of the electronic ground state can be extracted from the known values of the optical dielectric constant.

\begin{figure*}[h!tbp]
\includegraphics[width=.9\textwidth]{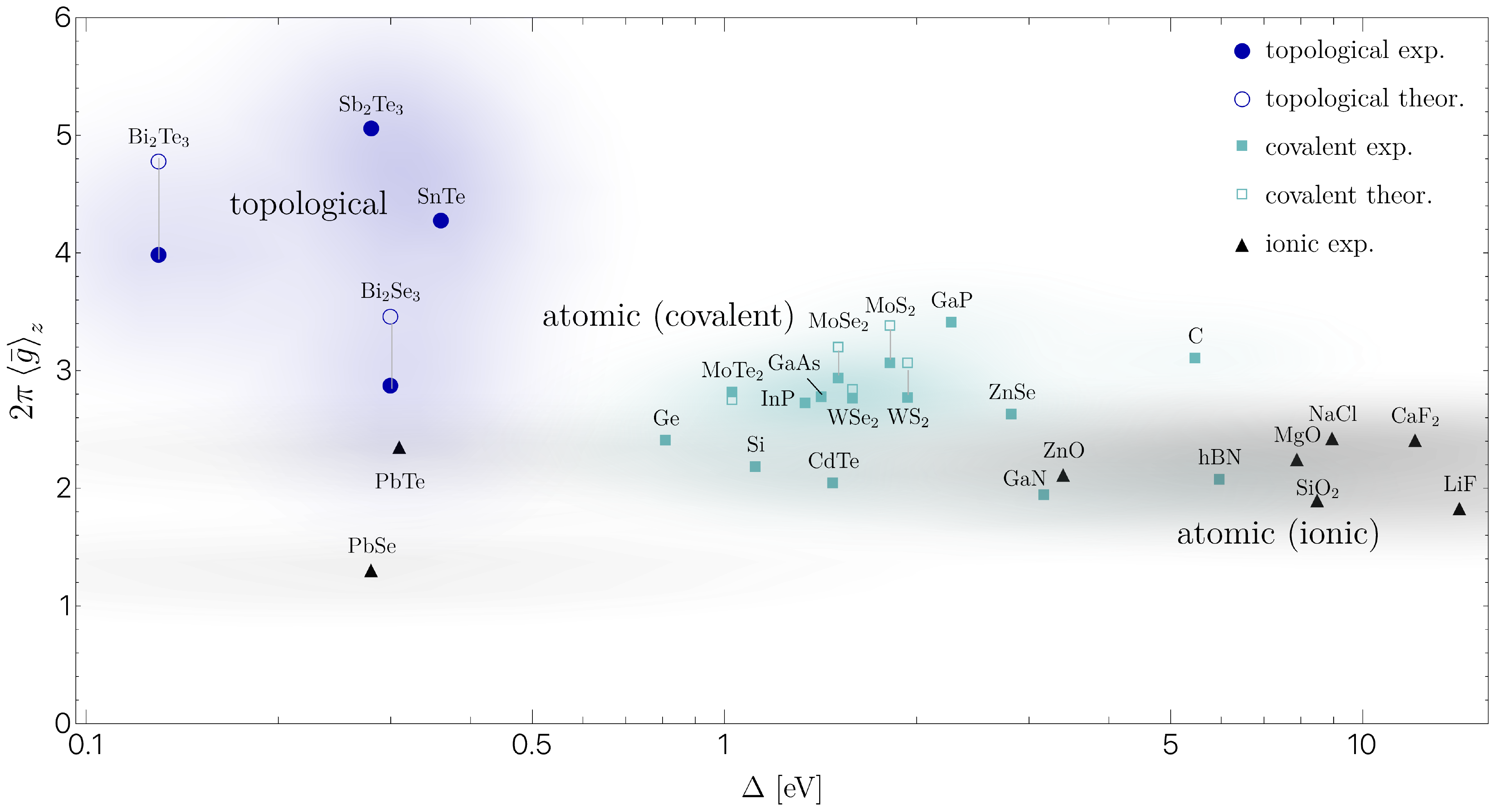}
\caption{Estimate of the quantum metric $\left< \bar g \right>_{z}$ given in \eqref{eq:chigdel} as a function of the spectral gap $\De$. The $\left< \bar g \right>_z$ values computed from the experimental data are represented by full markers, and the theoretical calculations from the fitted tight-binding models are shown with open symbols. One can identify three main groups, distinguished by color: On the right (black), with a gap above $4~\rm eV$ and small metric, are the atomic insulators with ionic bonding, where the electrons predominantly surround an atomic site. In the middle (green), with a gap in the range $\sim 1-2~\rm eV$ are covalent insulators where electrons live on the bonds. Examples include transition metal dichalcogenide monolayers, $sp^3$ bonded semiconductors, and also the monoatomic obstructed atomic limits $\rm Si$ and $\rm Ge$, where the electrons cannot move from the bond due to symmetry constraints. On the left (blue) are topological insulators with narrow gaps ($\sim0.2~\rm eV$). Insulators belonging to this group are characterized by extraordinary high refractive indices. Experimental values are according to~\cite{munkbat,LiuTMD,madelung, Kobayashi,Wing2021,hbcp,Bonzi.Grad.2022,Villars2016} and tabulated in the supplementary information, where we also comment on the fidelity of the estimate Eq.~\eqref{eq:chigdel} across various material classes}.
     \label{dielPlot}
\end{figure*}

To answer this question, we find it convenient to introduce a dimensionless localization marker that allows comparing materials with different geometries of the unit cell, and even different dimensionality. The quantity of interest is the out-of-plane average of the in-plane quantum metric, 
\beq
\label{eq:trgz}
\langle g\rangle_z =\int \frac{a_z dk_z}{2 \pi} \int \frac{d^2 \bfk}{(2\pi)^2} \, (g^{xx}+g^{yy}) \, ,
\eeq
where $a_z$ is the lattice constant in the $z$-direction. $2\pi\langle  g\rangle_z$ is dimensionless and can serve as a measure of localization for both two- and three-dimensional systems. The more the orbitals inside the material are spread in the $x$ and $y$ direction, the larger the value this quantity assumes.
Note that the choice of the $xy$-plane for the definition \eqref{eq:trgz} is an example, and the localization along the $z$-direction can be estimated analogously. For the sake of clarity, we restrict the analysis to the in-plane localization measure in materials with $C_3$ or $C_4$ rotational symmetry, where $\epsilon^{xx} = \epsilon^{yy} \equiv \epsilon$ holds.
From the combination of Eq.~\eqref{eq:bd} and Eq.~\eqref{eq:trgz}, one can infer an approximate relation between $\langle g\rangle_z$, the gap expressed in the units of energy, and the in-plane dielectric constant
\beq
\label{eq:chigdel}
{\langle g\rangle_z}\gtrsim \frac{ \epsilon_0}{e^2} a_z \De (\epsilon-1) \equiv \braket{\bar g}_z.
\eeq 
By construction, in the limit when the band gap $\Delta$ is large and the bands below and above the gap are nearly dispersionless, as occurs in the case of Landau levels, ${\langle g\rangle_z} \simeq \braket{\bar g}_z$ holds. Away from this limit, the estimate \eqref{eq:chigdel} is useful as a lower bound. A tighter estimate of the metric can be obtained by using the average gap in Eq.\eqref{eq:chigdel}. For a subset of materials presented in the supplementary information \cite{SM}, this estimate gives numerical values within 10\% of the ones obtained through tight-binding calculations.

In Fig.~\ref{dielPlot}, we display the values of $\left< \bar g \right>_z$ calculated from experimental values of $\epsilon$ and $\Delta$ (filled symbols) for a few topological and trivial materials  with band gaps ranging from a few hundred meV to several eV.
Materials with large gaps, on the right of Fig.~\ref{dielPlot}, are the atomic insulators with ionic bonding, whose electrons are well-bound to their original atoms: they show the lowest values of  $\langle\bar g\rangle_z$. In the middle of Fig. \ref{dielPlot}, we predominantly find covalent semiconductors, whose electrons live predominantly on the bonds. These include obstructed atomic limits (OALs)~\cite{TQC}, where symmetry fixes the center of the electronic cloud in a high symmetry point such as the bond center. In these cases we find $\langle\bar g\rangle_z$ to be consistently higher than in ionic insulators. Intriguingly, a large number of transition metal dichalcogenides (TMDs), which are OALs~\cite{TMDOAL}, possess a quite similar $2\pi\langle\bar g\rangle_z  \approx
2.8-3.0$ obtained from experimental values of $\epsilon$ and $\Delta$. 
These values agree reasonably well with theoretical estimates of Eq.~\eqref{eq:chigdel} when using tight-binding models to calculate $\epsilon$~\cite{LiuTMD} (empty symbols), implying that their capacitance is not affected much by high energy bands and core orbitals.
On the left side of Fig.~\ref{dielPlot}, at small energy gaps, we find strong topological insulators such as $\rm Bi_2 Se_3$, which should be contrasted against the trivial rocksalt semiconductors PbTe and PbSe with much lower $\langle\bar g\rangle_z$. 

Many topological insulators possess large values of $\langle\bar g\rangle_z$ and small values of $\Delta$, both contributing to exceptionally high dielectric constants, makng them ideal candidates for large refractive index materials. However, note that the enhancement of the values of $\langle\bar g\rangle_z$ in the topological insulator group is aided by the smallness of their spectral gaps, which puts these materials closer to the metal-insulator transition, and not only due to their topological nature. Finally, we point out the high value of $\langle\bar g\rangle_z$ occurring in the topological crystalline insulator SnTe, a material characterized by non-trivial mirror Chern number along certain high symmetry planes in momentum space \cite{Hsieh2012}.

An interesting outlier on the right side of the chart is diamond ($\rm C$), known to have a large gap and expected to lie at the bottom in Fig. \ref{dielPlot}. Nonetheless, an atomic obstruction prevents electrons in this material from localizing in the atomic sites forcing them to be pinned by symmetry at the bonds, which leads to an abnormally high $\langle\bar g\rangle_z$ compared to materials with similar gaps. For example, we can contrast diamond with $\rm GaN$, whose Wannier centers are situated close to the more electronegative element. The topological obstruction in diamond makes it a unique material with an exceptionally large gap and high refractive index, which leads to its unique brilliance emanating from trapped refracted light.

\section*{Discussion}
We have explored the intimate connection between the quantum geometry and the low-frequency behavior of the longitudinal conductivity provided by Eq.~\eqref{KuboTimeQGT}. Our analysis shows that in insulators, not only the polarization but also the conductivity is completely determined by the properties of the Hilbert space, as long as the time evolution of the quantum geometric tensor is taken into account.

Based on this finding, we have shown in several examples how an estimate of the ground state quantum metric can be obtained by measuring the intrinsic capacitance. 
Specifically for TBG aligned with hBN, as a function of the twist angle, we predict a sharp drop of the capacitance exactly at the magic angle, which is related to the corresponding decrease in the quantum metric.
Compared to other measurement schemes for the quantum metric using the excitation rate of on-shell electronic transitions via sum rules~\cite{Ozawa2018, Ozawa2018a, Tan2019, Yu2019a}, the approach presented here does not require access to a wide frequency range.

Historically, the precise relation between the dielectric constant and the gap size in insulators has remained unclear, despite intense efforts~\cite{Moss1985,Ravichandran2016,Gomaa2021,VanMechelen2022}. 
In light of this, the significance of Fig.~\ref{dielPlot} for the characterization of the dielectric properties of insulators is hard to overstate.
As explained in detail, guided by our results for the quasistatic conductivity, we suggest the rescaling by the out-of-plane lattice constant before attempting a comparison of $\e$ across materials, and conjecture that the remaining differences between materials with the same band gap but different $\epsilon$ are mostly due to the averaged quantum metric. While the relation presented in Eq.~\eqref{eq:chigdel} clearly needs to be studied for more example cases, these assumptions seem to work well for bulk and layered $3D$ materials.
Furthermore, we expect this approach to be useful in the search for new topological insulators, or for high refractive index materials.

A similar reformulation as the one demonstrated here might be possible for higher-order response functions, which can serve to illuminate the physical origin of the recently demonstrated corrections to the quantum anomalous Hall effect~\cite{Kaplan2023a}.

\begin{acknowledgments}
We thank 
N.~Regnault,
C.~Felser,
J.~Mitscherling,
E.~Khalaf and
J.~S.~Hofmann
for helpful discussions. IK sincerely acknowledges A. Pertsova for sharing tight-binding codes.
This work was supported by the NSF MRSEC program at Columbia through the Center for Precision-Assembled Quantum Materials (DMR-2011738). T.H.\ acknowledges financial support by the 
European Research Council (ERC) under grant QuantumCUSP
(Grant Agreement No. 101077020).
\end{acknowledgments}

\clearpage

\bibliography{refs,literature}

\pagebreak
\onecolumngrid
\newpage
\begin{center}
\textbf{\large Supplementary information for\\ ``The quantum geometric origin of capacitance in insulators"}
\\[6pt]
Ilia Komissarov, Tobias Holder, Raquel Queiroz
\end{center}

\renewcommand{\thefigure}{\arabic{figure}}
\renewcommand{\figurename}{Supplementary Figure}
\renewcommand{\tablename}{Supplementary Table}
\setcounter{figure}{0}

In this supplementary information, we present a detailed derivation of the relation between interband and intraband conductivity (Sec.~\ref{AppA}), quantum geometry (Sec.~\ref{AppB}), and the  dynamical polarization (Sec.~\ref{AppC}). We apply the obtained relations to various example systems: namely, we present in detail how the time-dependent polarization enters in Landau levels (Sec.~\ref{AppD}), derive the quantum geometric tensor for Landau levels (Sec.~\ref{AppE}), and develop some intuition behind these findings by comparing them with a semiclassical point of view (Sec.~\ref{AppF}). This is followed by a derivation of the capacitance in a gapped Dirac cone model (Sec.~\ref{AppG}). Lastly, we discuss the applications to the dielectric constant and present the complete data which was used to generate Figure~4 of the main text (Sec.~\ref{appendiel}).

\section{Interband and intraband conductivity}

\label{AppA}

In this section, we demonstrate how the Kubo formula for conductivity \cite{Kubo} can be split into intraband and interband terms. The former describes the conventional dissipative Fermi-surface transport, whereas the latter arises due to transitions between different bands. The starting point is the standard expression for the conductivity tensor (cf. e.~g.~\cite{Bradlyn2012}):
\beq
\label{kubof1}
\sigma^{\mu \nu}  (\omega) = \frac{i \bar n e^2}{m \omega_+} \delta^{\mu \nu} + \frac{1}{\hbar \omega_+ A} \int_0^\infty dt \, e^{ i \omega_+ t} \braket{[\hat J^\mu(t),\hat J^\nu(0)]} \, ,
\eeq
where $\bar n$ is the total charge carrier density, $m$ and $e$ are the mass and the charge of the carriers, and $A$ is the area of the conducting sample. The brackets $\braket{\ldots}$ denote the vacuum expectation value. The convergence of the integral is ensured by an infinitesimal relaxation rate, i.~e. $\w_+ = \w+ i \varepsilon$, $\varepsilon>0$. Henceforth, we mostly omit the subscript $+$, restoring it only where necessary to prevent singular behavior.

By inserting the complete basis of energy eigenstates $\ket{m \bfk}$, we write the commutator in Eq.~\eqref{kubof1} as
\beq
\label{Jcomm}
\braket{[\hat J^\mu(t), \hat J^\nu(0)]} = \sum \limits_{n m} \int \frac{A \, d^2 \bfk}{(2 \pi)^2} f_{nm}(\bfk)  \braket{n \bfk | \hat J^\mu | m \bfk} \braket{m \bfk| \hat J^\nu | n \bfk} e^{ i \omega_{nm}(\bfk) t}\, ,
\eeq
where the indices $m$ and $n$ enumerate the bands, $\w_{nm} \equiv \omega_n - \omega_m$, and $f_{nm} \equiv f_n-f_m$, where $f_n(\bfk) = \theta(E_F - \hbar \omega_n(\bfk))$ is the zero temperature Fermi-Dirac distribution function with respect to the Fermi energy $E_F$. 
The matrix elements of the current operators are evaluated in the basis of Bloch states $\ket{n\bfk} = u_n(\bfk) \ket{\bfk}$, where $u_n(\bfk)$ are the eigenstates of the Hamiltonian in the momentum space: $\hat H(\bfk) u_n(\bfk) = \hbar \omega_n(\bfk) u_n(\bfk)$. In the following, we omit the momentum index and likewise use the shorthand notation of the d-dimensional integral
\beq
\int_{\rm BZ} \equiv \int \frac{d^d \bfk}{(2 \pi)^d} \, .
\eeq
Plugging \eqref{Jcomm} into (\ref{kubof1}) and evaluating the time-integral, we obtain
\beq
\sigma^{\mu \nu} (\omega)= -\frac{i}{\hbar} \sum\limits_{nm} \int_{\rm BZ} \frac{f_{nm}}{\omega_{nm}}  \frac{ J^\mu_{nm} J^\nu_{mn}}{ \omega_{nm} + \omega} \, ,
\label{KuboAllSum}
\eeq
where the first (diamagnetic) term in \eqref{kubof1} was used to subtract the singular in $\omega \to 0$ limit piece in the commutator term. The remaining sum in Eq.~\eqref{KuboAllSum} can be split into two contributions: The \textit{intraband} part with $m=n$ and the \textit{interband} part where $m \neq n$.
The intraband conductivity is obtained from (\ref{KuboAllSum}) by substituting
\beq\bal
f_{nm}& \to f \( E_F - E_n(\bfk + \bfq) \) - f\( E_F - E_n(\bfk) \)\,  ,\\
\omega_{nm} & \to \omega_n(\bfk + \bfq) - \omega_n(\bfk) \, ,
\eal\eeq
and taking a limit $\bfq \to 0$
\beq
\s_{\rm intraband}^{\mu \nu} (\omega)= \frac{i}{\omega} \sum\limits_{n} \int_{\rm BZ} f'_n   J^\mu_{nn}   J^\nu_{nn} \, .
\label{KuboIntra}
\eeq
At zero temperature, the derivative of the Fermi function is simply a delta function $\d(E_F - \hbar \omega_n)$ that selects all the points in the BZ where the bands cross the Fermi surface. Hence, this contribution vanishes in insulators in the absence of disorder.

The interband ($m\neq n$) contribution to the sum (\ref{KuboAllSum}) is convenient to express in terms of the position operators to make an explicit connection with quantum geometric quantities such as Berry curvature and quantum metric. In order to do so, we utilize
\beq
\bal
J^\mu_{nm} = \frac{e}{\hbar} \braket{n| \prl^{\mu} \hat H |m} & = \frac{e}{\hbar} \bra{n} \prl^{\mu} \( \hat H \ket{m}  \) -  \frac{e}{\hbar} \bra{n}  \hat H \ket{ \prl^{\mu} m} \\ & =  \frac{e}{\hbar} (E_m - E_n) \braket{n | \prl^{\mu} m} 
= - i e \omega_{n m} r_{n m}^\mu \, .
\eal
\label{see}
\eeq
Upon using $J^{\mu}_{nm} = - i e \omega_{nm}  r^{\mu}_{nm}$, the interband contribution becomes
\beq
\sigma^{\mu \nu}_{\rm interband} (\omega)= - \frac{i e^2}{\hbar} \sum\limits_{n \neq m} \int_{\rm BZ} f_{nm} \omega_{nm}  \frac{ r^\mu_{nm} r^\nu_{mn}}{ \omega_{nm}+\omega} \, .
\label{KuboInter}
\eeq

The above term does not vanish in insulators even at zero frequency. For example, in the limit $\omega \to 0$ the expression above takes the form of the celebrated TKNN formula \cite{TKNN}
\beq
\bal
 \sigma^{\mu \nu}_{\rm interband}(0) & = - \frac{i e^2}{\hbar} \sum\limits_{n \neq m} \int_{\rm BZ} (f_n - f_m)  r^\mu_{nm} r^\nu_{mn} = - \frac{i e^2}{\hbar} \sum\limits_{n, m} \int_{\rm BZ} f_n ( r^\mu_{nm} r^\nu_{mn} -  r^\nu_{nm}  r^\mu_{mn} ) \\
 & = - \frac{i e^2}{\hbar} \sum\limits_{n} \int_{\rm BZ} f_n ( \braket{n| \hat r^\mu \hat r^\nu |n} - \braket{n| \hat r^\nu \hat r^\mu |n} ) = - \frac{i e^2}{\hbar} \sum\limits_{n} \int_{\rm BZ} f_n (\braket{ \prl^\mu n| \prl^\nu n} - \braket{\prl^\nu n| \prl^\mu n} ) \, ,
\eal
\eeq
i.e. the conductivity is, up to a constant, the sum of the Chern numbers of the occupied bands. At non-zero frequency, on the other hand, it is not obvious that the interband contribution (\ref{KuboInter}) is geometric in its origin: something one would expect to be the case in an insulator, at least in the small $\omega$ regime. To investigate this, one may further write:
\beq
\bal
\label{interfnmmn}
\sigma^{\mu \nu}_{\rm interband} (\omega)= -  \frac{i e^2}{\hbar} \sum\limits_{n \neq m} \int_{\rm BZ} & \left[ f_{n}(1-f_m)-f_{m}(1-f_n) \right] \,\omega_{nm}  \frac{ r^\mu_{nm}  r^\nu_{mn}}{\omega_{nm} + \omega}\\ & = - \frac{i e^2}{\hbar} \sum\limits_{n \neq m} \int_{\rm BZ} f_n(1-f_m) \, \omega_{nm} \left( \frac{  r^\mu_{nm}  r^\nu_{mn}}{\omega_{nm} + \omega} - \frac{r^\nu_{nm}  r^\mu_{mn}}{\omega_{nm}-\omega} \right) \, ,
\eal
\eeq
where we relabeled the summation indices in the second line. One may proceed by introducing the interband matrix elements of the Berry curvature and the quantum metric
\beq
\Omega_{nm}^{\mu \nu} = i ( r^\mu_{nm}  r^\nu_{mn} -  r^\nu_{nm}  r^\mu_{mn} ) \, , \qquad g_{nm}^{\mu \nu} = \half( r^\mu_{nm}  r^\nu_{mn} +  r^\nu_{nm}  r^\mu_{mn})\, .
\eeq
With the definitions above, interband conductivity reads
\beq
\label{InterMetrCurv}
\sigma^{\mu \nu}_{\rm interband} (\omega)=  \frac{2 i e^2}{\hbar} \sum\limits_{n \neq m} \int_{\rm BZ} f_n(1-f_m) \frac{\omega \omega_{mn}}{\omega_{mn}^2-\omega^2} g^{\mu \nu}_{n m} - \frac{ e^2}{\hbar} \sum\limits_{n \neq m} \int_{\rm BZ} f_n(1-f_m) \frac{\omega_{mn}^2}{\omega_{mn}^2-\omega^2} \Omega^{\mu \nu}_{n m} \, .
\eeq
Note that the second term never contributes to the longitudinal conductivity, since $\Omega_{nm}^{\mu \nu}$ is an antisymmetric tensor. The metric $g_{nm}^{\mu \nu}$, on the other hand, may have off-diagonal components and contribute to $\sigma^{xy}(\omega)$. As we can see, the dispersion-dependent factors in the expression above prevent us from re-summing all the interband transitions into the ground state quantities: Berry curvature and quantum metric
\beq
\Omega^{\mu \nu} = i \sum\limits_{n \neq m} f_n(1-f_m) ( r^\mu_{nm}  r^\nu_{mn} -  r^\nu_{nm}  r^\mu_{mn} ) \, , \qquad g^{\mu \nu} =\frac{1}{2} \sum\limits_{n \neq m} f_n(1-f_m) \( r^\mu_{nm}  r^\nu_{mn} +  r^\nu_{nm}  r^\mu_{mn}\)\, .
\eeq
This is a consequence of the non-adiabaticity introduced by the presence of $\omega$. Nevertheless, it does not mean that the interband contribution (\ref{KuboInter}) is not purely quantum-geometric. As we will show explicitly in the next section, by considering the geometry of wavefunctions in space-time ($\bfk$, $t$), one is able to express \eqref{KuboInter} solely in terms of the Hilbert space quantities.

\section{Time-dependent quantum geometric tensor}
\label{AppB}

\begin{figure}[h!]
         \centering
         \includegraphics[width=0.33 \textwidth]{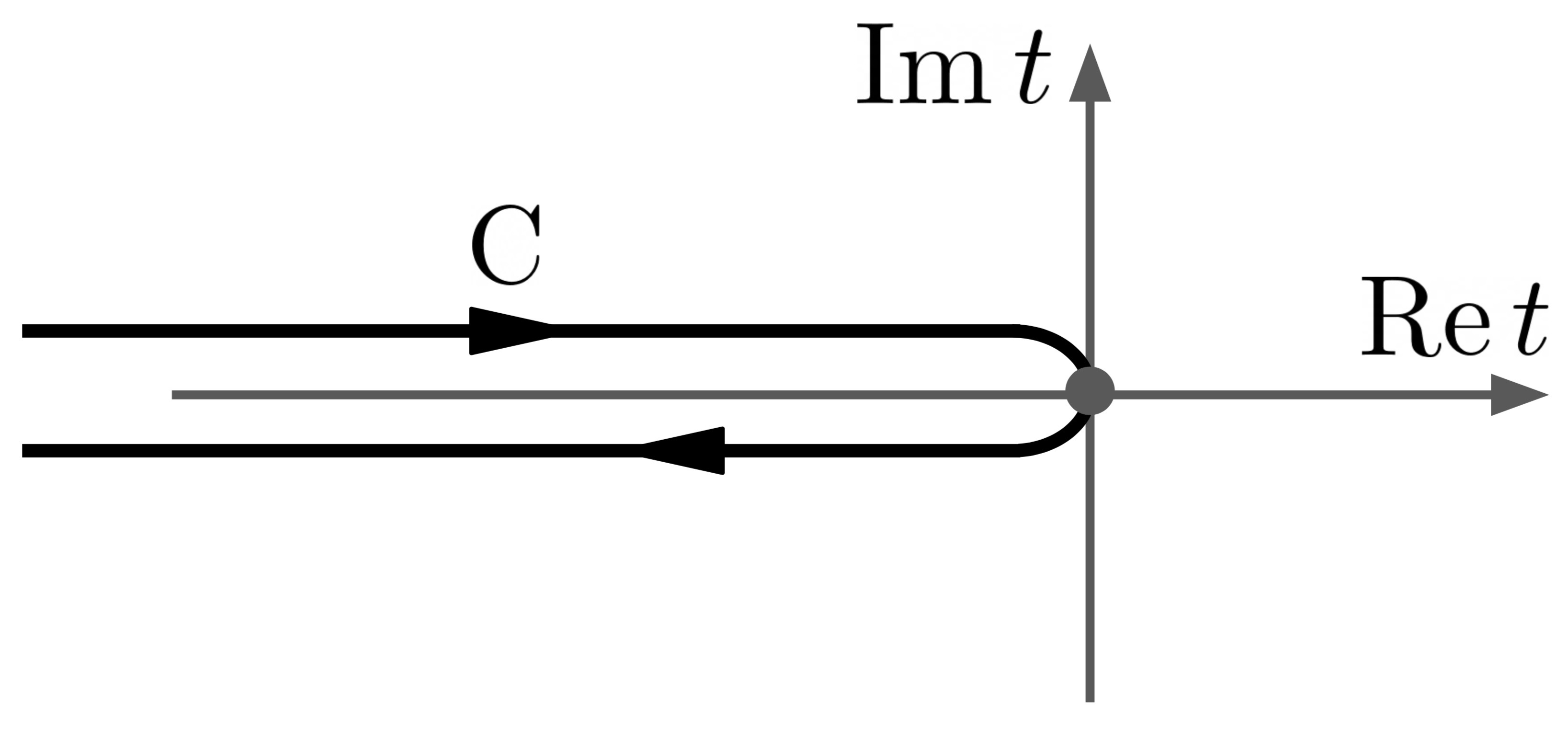}
         \caption{Keldysh contour in complex time. The point at the axes crossing corresponds to $t=0$.}
         \label{Keldysh}
     \end{figure}

In clean insulators, with a finite excitation gap, quasiparticle transport is impossible, and one would not expect the dispersion to enter the response functions. It implies that the factors of $\omega_{nm}$ in (\ref{KuboInter}) can be absorbed into the wavefunctions. Below we show that it is indeed the case and can be done by taking the time dependence of states (or operators) into account.
We start with the expression \eqref{interfnmmn}
\beq
\sigma^{\mu \nu}_{\rm interband} (\omega)= - \frac{i e^2}{\hbar} \sum\limits_{n \neq m} \int_{\rm BZ} f_n (1-f_m)\, \omega_{nm} \(  \frac{  r^\mu_{nm}  r^\nu_{mn}}{\omega_{nm} + \omega} - \frac{  r^\nu_{nm}  r^\mu_{mn}}{\omega_{nm} - \omega} \) \, .
\eeq
We further decompose
\beq
\frac{\omega_{nm}}{\omega_{nm} + \omega} = 1 - \frac{\omega}{\omega_{nm} + \omega} \, , \qquad \frac{\omega_{nm}}{\omega_{nm} - \omega} = 1 + \frac{\omega}{\omega_{nm} - \omega} \, ,
\eeq
which yields
\beq
\sigma^{\mu \nu}_{\rm interband} (\omega)= - \frac{e^2}{\hbar} \int_{\rm BZ} \Omega_{\mu \nu} + \frac{i e^2 \omega}{\hbar} \sum\limits_{n \neq m} \int_{\rm BZ} f_n (1-f_m) \(\frac{  r^\mu_{nm}  r^\nu_{mn} }{\omega_{nm}+
\omega}  -  \frac{  r^\nu_{nm}  r^\mu_{mn}}{\omega_{mn}+
\omega}    \) \, .
\label{KuboStep}
\eeq
Reinstating the infinitesimal imaginary part of the frequency $\omega \to \omega_+$ to enforce convergence, and using the Schwinger representation
\beq
\frac{i}{\omega_{nm} + \omega_+} = \int_{- \inf}^{0} d t \, e^{-i(\omega_{nm} + \omega_+) t} \, ,
\eeq
one can show that the terms with propagators in \eqref{KuboStep} can be conveniently wrapped up as
\begin{align}
&\quad  \frac{e^2 \omega}{\hbar} \sum\limits_{n \neq m}  \int_{\rm BZ}  f_n (1-f_m) \left( \int_{- \inf}^0 dt\, e^{-i (\omega_{nm}+\omega_{+} ) t}  r^\mu_{nm} r^\nu_{mn} - \int_{-\inf}^{0} dt \, e^{-i (\omega_{mn}+\omega_{+}) t}  r^\nu_{nm}  r^\mu_{mn}  \right) \\ 
&=  \frac{e^2 \omega}{\hbar} \sum\limits_{n \neq m} \int_{\rm BZ}  f_n (1-f_m) \left( \int_{- \inf}^0 dt\, e^{-i \omega_+ t} r^\mu_{nm}  r^\nu_{mn}(t)  + \int_{0}^{-\inf} dt \, e^{-i \omega_+ t}  r^\nu_{nm} (t)r^\mu_{mn} \right) \\  &  
=  \frac{e^2 \omega}{\hbar} \int_{\rm C} dt \int_{\rm BZ} e^{-i \omega_+ t} \, \hat T Q^{\mu \nu}(t) 
\, ,
\end{align}
where the integral is taken over the Keldysh contour C sketched in \autoref{Keldysh}, $\hat T$ denotes the (advanced) ordering of operators along $\rm C$, and we introduced the time-dependent ``quantum geometric tensor'' 
\beq
Q^{\mu \nu}(t) \equiv {\rm Tr} \lb \hat  P \hat r^\mu (0) (1-\hat P) \hat r^\nu (t) \rb \, ,
\eeq
where the operators appearing without the time label are taken at the initial time $t = 0$, and the trace runs over band indices. The interband contribution to the Kubo formula for conductivity, therefore, assumes the form 
\beq
\sigma^{\mu \nu}_{\rm interband} (\omega)= - \frac{e^2}{\hbar} \int_{\rm BZ}\( \Omega^{\mu \nu}  -  \omega \int_{\rm C} dt \, e^{- i \omega_+ t} \, \hat T Q^{\mu \nu}(t) \) = - i \frac{e^2}{\hbar} \int_{\rm BZ} \int_{\rm C} e^{- i \w_+ t} \frac{d}{d t} \left( \hat T Q^{\mu \nu}(t) \right)\, .
\label{KuboTimeQGT2}
\eeq

As seen from the above, the time-dependent quantity $Q^{\mu \nu}(t)$ solely determines the interband part of the conductivity, as a function of the time-dependent quantum geometry. For example, in the case of Landau levels, we have $Q^{\mu \nu}(t) =Q^{\mu \nu} e^{ i \omega_c t}$, and the momentum integral is taken over the reciprocal magnetic unit cell of the measure $2 \pi/l_{\rm B}^2$. Plugging these expressions into Eq.~\eqref{KuboTimeQGT2}, one obtains the ac conductivity tensor for Landau levels, Eq.~\eqref{LandauCond}. Hence, as proposed, the conductivity in insulators is entirely determined by the geometry of the Hilbert space.

Surprisingly, it turns out that not only the interband term allows for the geometric interpretation. In the next section, we take this reasoning one step further and show that the entire Kubo formula, including the intraband part, follows from the expectation value of the time-dependent position operator.

\section{Kubo formula for conductivity as a derivative of the polarization}
\label{AppC}

In this section, we demonstrate that the schematic expression inspired by the modern theory of polarization \cite{RestaModern}
\beq
\sigma^{\mu \nu} = \frac{\d}{\d E_\mu (t)} j^\nu(t) = \frac{\d}{\d E_\mu (t)} \frac{d}{d t} \left< P^\nu (t) \right>\, , 
\label{condAppShort}
\eeq
is not merely a convenient symbolic way of introducing electrical conductivity. On the other hand, quite literally, (\ref{condAppShort}) is the Kubo formula for conductivity. In order to show it, we write
\beq
\sigma^{\mu \nu} = \frac{\d}{\d E_\mu (t)} \frac{d}{d t} \left< \frac{- e  \hat r^\nu (t)}{A} \right> = - \frac{i \omega}{A} \frac{\d}{\d E_\mu (t)} \left< e  \hat r^\nu (t) \right> \, ,
\label{condApp}
\eeq
where the expectation value of the time-dependent position operator in the Heisenberg picture is obtained via the Schwinger-Keldysh formalism with the interaction Hamiltonian (note the second term required by hermiticity)
\beq
\label{Hint+-}
H_{\rm int} = J^\mu A_\mu = \frac{i}{\omega} J^\mu E_\mu e^{i \omega t} - \frac{i}{\omega} J^\mu E_\mu e^{-i \omega t} \, .
\eeq
The vacuum expectation value of the position operator at time $t$ is then
\beq
\label{keldyshEq}
\bal
 \frac{1}{A} \sum\limits_n f_n\braket{n |  e  \hat r^\nu (t) | n} = \sum\limits_n \int_{\rm BZ} f_n & \bra{n}  \( 1 - \frac{1}{\hbar \omega} \int_{- \inf}^t \D t'\, e^{i \omega t'} \hat J^\mu_I(t') E_\mu \) (e \hat r^\nu_I(t)) \times \\ & \qquad\qquad\qquad  \times \( 1 - \frac{1}{\hbar \omega} \int_{- \inf}^t \D t'\, e^{i \omega t'} \hat J^\sigma_I(t') E_\sigma \) \ket{ n } + {\rm c.c.} + O(E^2) \, ,
\eal
\eeq
where the subscript $I$ stands for the interaction picture with operators enjoying the ``unperturbed'' time-dependence $\hat A_{mn, I}(t)=e^{-i \omega_{mn} t} \hat A_I(0)$: in the following, to shorten the notation, we drop the subscript $I$. The expression c.c. stands for ``complex conjugate'' and contains the group of terms that oscillate with the frequency $e^{-i \omega t}$, which will be eliminated by the functional derivative $\d/\d E(t)$.

The $O(E^0)$ term in the expression (\ref{keldyshEq}) is related to the net Berry phase of the occupied bands
\beq
 i e \hbar \sum\limits_n \int_{\rm BZ} f_n  \braket{n|\prl^\mu n} \, .
\eeq
It corresponds to the ferroelectric polarization that material possesses in the absence of an external electric field.
The remaining terms represent the dielectric response: the fluctuation of the electric dipole moment under the influence of the external field. We further omit the $O(E^2)$ terms that contain the information about the non-linear response: whether they reproduce the known non-linear contributions to conductivity is an interesting question that we leave for future work. 

 Keeping only the terms linear in $E_\nu$, inserting another sum over energy eigenstates $\sum_n \ket{m} \bra{m}$, and evaluating the time integrals, we write
\begin{align}
\frac{1}{A} \sum\limits_n f_n\braket{n |  e  \hat r^\nu (t) | n} = & - \sum_{mn}\int_{\rm BZ} f_n \( \frac{1}{\hbar \omega} \frac{e^{i (\omega + \omega_{nm})t}}{i (\omega_{nm}+\omega)} \braket{n|\hat J^\mu(0)|m} E_\mu \)\braket{m|e \hat r^\nu(0)|n} e^{-i \omega_{nm} t} \\
& -\sum_{mn}\int_{\rm BZ} f_n \braket{n|e \hat r^\nu(0)|m} e^{i \omega_{nm} t} \(\frac{1}{\hbar \omega} \frac{e^{i (\omega - \omega_{nm})t}}{i (\omega - \omega_{nm})} \braket{m|\hat J^\mu(0)|n} E_\mu\) + \ldots
\end{align}
The time dependence in this expression is canceled by the functional derivative, and we obtain
\beq
\frac{1}{A} \( \frac{\d}{\d E_\mu (t)} \braket{ e \hat r^\nu(t)} \) = \frac{i e}{\hbar \omega} \sum\limits_{mn} \int_{\rm BZ} \, f_n \(\frac{ J^\mu_{nm}  r^\nu_{mn} }{\omega_{nm}+\omega} - \frac{ r^\nu_{nm} J^\mu_{mn}}{\omega_{nm} -  \omega}\) \, .
\eeq
We then interchange the summation indices in the second bracket:
\beq
-\frac{i \omega}{ A} \( \frac{\d}{\d E_\mu}\braket{e \hat r^\nu(t)} \) =  \frac{e}{\hbar} \sum\limits_{mn} \int_{\rm BZ} \, \(f_n - f_m \) \frac{ J^\mu_{nm}  r^\nu_{mn}}{\omega_{nm} + \omega} \, .
\label{KuboPosVel}
\eeq
As soon as $m$ and $n$ are different, we are formally allowed to plug in $J^\mu_{nm} \to - i e \omega_{nm} r^\mu_{nm}$, so we arrive at the interband piece of the Kubo formula (\ref{KuboInter}) :
\beq
\sigma^{\mu \nu}_{\rm interband} (\omega)=- \frac{i e^2}{\hbar} \sum\limits_{n \neq m} \int_{\rm BZ} f_{nm} \omega_{nm}  \frac{  r^\mu_{nm}  r^\nu_{mn}}{\omega_{nm} + \omega} \, .
\label{kk1}
\eeq
In the case $m$ and $n$ are equal, we use the standard trick of splitting the initial and final states' momenta $\ket{n} \to \ket{n\bfk'}$, $\ket{m} \to \ket{n\bfk}$ and use the identity that follows from the distributional properties of the delta function
\beq
\braket{n\bfk' | \hat r^\mu |n \bfk} = - i \hbar \d (\bfk-\bfk') \(u_n^\dagger(\bfk) \prl^\mu u_n(\bfk)\) + i \hbar \prl^\mu \d(\bfk- \bfk') \, .
\eeq
As we plug this expression into (\ref{KuboPosVel}), the contribution from the first term containing a delta function vanishes after the momentum integration as $f_m \to f_n$. The the second term can be evaluated using integration by parts, which shifts the momentum derivative from the delta function to the Fermi distribution bringing about another factor of velocity. The resulting expression is nothing but the Drude intraband term (\ref{KuboIntra})
\beq
\sigma_{\rm intraband}^{\mu \nu} (\omega)=  \frac{i}{\omega} \sum\limits_{n} \int_{\rm BZ} f_n'  J^\mu_{nn} J^\nu_{nn} \, .
\label{k31}
\eeq
Hence, Eq.~\eqref{condAppShort} is indeed the Kubo formula for conductivity.
 
 \section{Relation between time-dependent polarization and conductivity in Landau levels}
 \label{AppD}
  In this section, we illustrate how electrical conductivity can be obtained directly from the time-dependent electric dipole moment via the relation
 \beq
 \sigma^{\mu \nu} = - \frac{i \omega}{A} \frac{\d}{\d E_\mu (t)} \left< e  \hat r^\nu (t) \right> \, .
 \label{LLder}
 \eeq
 We perform this check for the Landau problem in oscillating in-plane electric field $E_x(t)=E e^{i \omega t}$, where the matrix element $\braket{ \hat r^\nu (t)}$ can be explicitly obtained.

It is convenient to choose the $y$-translationally-invariant gauge 
\beq
A_x = \frac{i}{\omega}\E e^{ i \omega t} \, , \qquad A_y = B x\, ,
\eeq
such that $p_y$ is conserved. The time-dependent Schrödinger equation then assumes the form
\beq
\label{SchEq}
\lb \frac{(\hat p_x + \frac{i e}{\omega} E e^{ i \omega t} + \rm{c.c.} )^2}{2m_e} + \frac{(\hat p_y + e B x)^2}{2 m_e} \rb \psi(x, y, t) = i \hbar \, \prl_t \psi(x,y,t) \, .
\eeq
One of the solutions is given by the gaussian ansatz
\beq
\psi_{p_y}(x, y, t) = {\cal N}(t) e^{i p_y y/\hbar} e^{-i \vp(t)} \exp \lb -\frac{(x-x_c(t))^2}{2 l_B^2} \rb\, ,
\label{ansatz}
\eeq
where $\vp(t)$ is an energy phase inessential for the discussion, and
\beq
\bal
x_c(t) =& -\frac{p_y}{m_e \omega_c} +x_1(t)+i x_2(t)\, , 
\qquad &
x_1(t)  &= - \frac{2 e E}{m_e}\frac{\cos(\omega t) }{\omega_c^2 - \omega^2}\, , \\  x_2(t)   =& \frac{2 e E}{m_e} \frac{\omega_c}{\omega} \frac{ \sin(\omega t) }{\omega_c^2 - \omega^2} \, , & {\cal N}(t) &= \frac{1}{\sqrt{\sqrt{\pi}l_{B}}} e^{-x_2^2(t)/2 l_B^2} \, .
\eal
\eeq
The wave function Eq.~\eqref{ansatz} should be understood as a time-dependent analog of the lowest Landau level. We then proceed to explicitly compute the matrix element of the position operator $\hat x$:
\beq
\bal
\braket{\psi_{p_y}|\hat x|\psi_{p_y}} & =  \frac{1}{\sqrt{\pi} l_B} \int dx \, x \exp \lb - \frac{(x-x_1(t)+p_y/m_e \omega_c)^2}{l_B^2} \rb = \\
& = x_1(t) - p_y/m_e \omega_c = - \frac{e E}{m_e} \frac{1}{\omega_c^2-\omega^2}(e^{i \omega t} + e^{-i\omega t}) - p_y/m_e \omega_c\, .
\eal
\eeq
Taking the functional derivative (\ref{LLder}), for the fully occupied Landau level ($A/N=\Phi_0/B=h/eB$) we obtain
\beq
\sigma^{xx}  = i \frac{e^2}{h} \frac{\omega \omega_c}{\omega_c^2 - \omega^2}\, ,
\eeq
in agreement with the result obtained from the Kubo formula in~Appendix \ref{AppE}. Remarkably, all we needed to extract this result is the expectation value of the position operator at the time $t$.

Unfortunately, our choice of gauge does not allow for the calculation of $\braket{\hat y(t)}$, since the wavefunction Eq.~\eqref{ansatz} does not decay in the $y$-direction. For completeness, however, we proceed by evaluating the current $j^y$, which is always well-defined: 
\beq
\bal
j^y & =  -\frac{e}{m_e} \int_{- \inf}^{\inf} \frac{dp_y}{2 \pi \hbar} \psi_{p_y}^*(x,t)(p_y + e B x)\psi_{p_y}(x,t) \\ &  =  -\frac{e}{m_e} \int_{- \inf}^{\inf} \frac{dp_y}{2 \pi \hbar} \psi_{p_y}^*(x,t) (m_e \omega_c x_1(t)) \psi_{p_y}(x,t) =2 \frac{e^2}{h} \frac{\omega_c^2}{\omega_c^2 - \omega^2} E \cos(\omega t)\, ,
\eal
\eeq
from which we obtain
\beq
\sigma^{\mu \nu} = \frac{\d j^\mu}{\d E_\nu(t)} ~ \longrightarrow ~ \sigma^{xy}(\omega) = - \frac{e^2}{h} \frac{\omega_c^2}{\omega_c^2 - \omega^2} \, .
\eeq
In the stationary limit, this expression reduces to the well-known value for a fully occupied Landau level $\s_{xy} = - e^2/h$.

\section{Quantum geometry and conductivity in Landau levels}

\label{AppE}

Consider the Hamiltonian for a two-dimensional electron gas in a uniform magnetic field
\beq
\hat H_{\rm LL} = \frac{\hat \pi^\mu \hat \pi_{\mu}}{2 m_e}\, , \qquad \hat \pi^\mu = \hat p^\mu + e \hat A^{\mu} \, , \qquad B = \prl^x \hat A^y - \prl^y \hat A^x = {\rm const} \, ,
\eeq
where $\hat p^\mu$ are the canonical momenta. The kinematical momenta $\hat \pi^\mu$, up to a normalization constant, commute canonically
\beq
[\hat \pi^\mu, \hat \pi^\nu] = - i \hbar e B \varepsilon^{\mu \nu}\, , \qquad  \varepsilon^{\mu \nu} = \begin{pmatrix}
    0 & 1 \\
    -1 & 0
\end{pmatrix} \, ,
\eeq
which allows to define the interlevel ladder operators $a$, $a^\dagger$ with $a = (\pi^x - i \pi^y)/\sqrt{2 \hbar e B}$. In order to compute quantum geometric quantities, it is convenient to have matrix elements of $\hat r^\mu$ to be well-defined. Hence, we adopt the radial gauge $A^\mu=- \varepsilon^{\mu \nu} x^\nu B/2$.
It is then useful to define another set of momenta 
\beq
\hat{\tilde \pi}^\mu = \hat p^\mu - e \hat A^\mu\, , \qquad [\hat{ \tilde \pi}^\mu, \hat {\tilde \pi}^\nu] = i \hbar e B\varepsilon^{\mu \nu}\, ,
\eeq
that commute with any $\pi^\mu$, and hence can be used to define ladder operators responsible for the degeneracy of the Landau levels. Using the definitions of $\pi^\mu$ and $\tilde \pi^\mu$, we express the coordinate operators in terms of the momenta
\beq
\hat r^\mu = \frac{l_B^2}{\hbar}  \varepsilon^{\mu \nu}(\hat{ \pi}^\nu - \hat {\tilde \pi}^\nu)\, .
\eeq
The quantity we are interested in is the quantum geometric tensor (QGT) defined as
\beq
Q^{\mu \nu} = {\rm Tr} \lb \hat P \hat r^\mu \(1-\hat P\) \hat r^\nu \rb = \braket{n-1| \hat r^\mu | n}\braket{n| \hat r^\nu | n-1}\, ,
\label{QGTlandau}
\eeq
where we accounted for the fact that only matrix elements taken between neighboring Landau levels are non-zero. Real and imaginary parts of the QGT are the quantum metric and Berry curvature
\beq
Q^{\mu \nu} = g^{\mu \nu} - \frac{i}{2}\Omega^{\mu \nu}\, .
\eeq

Since the matrix element of $\hat{\tilde \pi}^\nu$ vanishes between states with the same momentum, we plug $\hat \pi^x = \sqrt{\hbar e B} (\hat a+\hat a^\dagger)/\sqrt{2}$, $\hat \pi^y = i \sqrt{\hbar e B} (\hat a-\hat a^\dagger)/\sqrt{2}$ into Eq.~\eqref{QGTlandau}, and evaluate
\beq
\bal
Q^{\mu\mu} & = g^{\mu\mu} = \frac{l_B^2}{2}C \, , \\
Q^{xy} & = -\frac{i}{2}\Omega^{xy} = - i \frac{l_B^2}{2} C \, ,
\eal
\eeq
where we identified the number of occupied Landau levels $n-1$ with a Chern number $C$.
To determine conductivity, we use the expression for the conductivity tensor, Eq.~\eqref{InterMetrCurv} via the metric and Berry curvature, and utilize $\omega_{nm} = \d_{n-1,m} \omega_{\rm c}$, from which we immediately obtain the conductivity matrix
\beq
\sigma^{\mu \nu} = \frac{e^2}{h} C \frac{\omega_c^2}{\omega_c^2 - \omega^2} \begin{pmatrix}
    i \omega/\omega_c & -1 \\
    1 & i \omega/\omega_c
\end{pmatrix} \, ,
\label{LandauCond}
\eeq
where $\omega_c=eB/m_e$ is the cyclotron frequency. Inverting the above tensor, we find the resistivity matrix
\beq
\r^{\mu \nu} = \frac{h}{e^2 C} \begin{pmatrix}
    i \omega/\omega_c & 1 \\
    -1 & i \omega/\omega_c
\end{pmatrix} \, .
\label{LandauRes}
\eeq

\section{Classical derivation of capacitance in Landau levels}
\label{AppF}
It is well-known that many classical and quantum results obtained for the harmonic oscillator (and so, Landau levels) take the same form, and, as we will see, the result for the capacitance is no exception.
In the following we briefly discuss this classical derivation of the longitudinal conductivity of the 2DEG in order to highlight the physical origin of this effect.

We consider classical free-electron gas in two dimensions subject to a perpendicular magnetic field as well as a harmonically oscillating electric force directed along the $x$-axis. The equation of motion of the individual electrons is
\beq
\ddot x(t) + \omega_{\rm c}^2 x(t) = - \frac{e}{m_e} E e^{i \omega t}\, .
\eeq
The above equation is solved by the time-dependence
\beq
x(t) = -\frac{e E e^{i \omega t}}{m_e(\omega_{\rm c}^2 - \omega^2)} \, .
\eeq
The $x$-polarization of the system is obtained as a net dipole moment per unit area
\beq
P(t) = \frac{N}{A} (-e x(t)) = \frac{N}{A} \frac{e^2}{m_e(\omega_c^2-\omega^2)} E e^{ i \omega t}\, .
\eeq
The longitudinal conductivity is defined as
\beq
\label{sxxClass}
\sigma^{xx} = \frac{\d}{\d E(t)} \frac{d}{d t} P(t) = i \omega \frac{N}{A}\frac{e^2}{m_e (\omega_c^2 - \omega^2)}\, .
\eeq
To make a connection with the quantum result [Eq.~\eqref{LandauCond}], we assume that the number of electrons $N$ is just enough to occupy $C$ Landau levels, i.e. $N \Phi_0/(A B) = C$, where $\Phi_0 = h/e$---flux quantum. Plugging $N$ into (\ref{sxxClass}), we obtain
\beq
\sigma^{xx} = i\frac{e^2 C}{h} \frac{\omega \omega_{\rm c}}{\omega_c^2-\omega^2}\, ,
\eeq
in full consistency with $\s_{xx}$ found in Eq.~\eqref{LandauCond}.

\section{$2D$ Gapped Dirac cone and Haldane model}

\label{AppG}

In this section, we discuss the capacitive conductivity of a two-dimensional gapped Dirac cone dispersion and its relation to the topology. We begin by considering the following Hamiltonian: 
\beq
\label{coneham}
\hat H = k_x \hat \s_x + k_y \hat \s_y + M \hat \s_z \, .
\eeq
The spectrum and the normalized eigenfunctions in this model are
\beq
E^{\pm} = \pm \sqrt{k^2+M^2} \, , \qquad u^{\pm}(\bfk) = \sqrt{\frac{E^+ \mp M}{2 \omega^+}} \( \frac{M \pm E^+}{k_x + i k_y}\, ,~~ 1 \)^T \, , \qquad k = \sqrt{k_x^2+k_y^2} \, .
\eeq
The diagonal components of the quantum metric can be found from $g^{\mu\mu} = \left|\left<u^+|\prl^\mu u^-\right>\right|^2$, which gives
\beq
g^{xx} = \frac{k_y^2 + M^2}{4(k^2 + M^2)^2}\, , \quad g^{yy} = \frac{k_x^2 + M^2}{4(k^2 + M^2)^2}\, , \quad  g \equiv g^{xx}+g^{yy}=\frac{k^2 + 2 M^2}{4(k^2 + M^2)^2}\, . 
\label{conemetric}
\eeq
Using the isotropy of the model (\ref{coneham}), it is convenient to express the longitudinal conductivity as
\beq
\sigma^{xx} = \frac{\sigma^{x x} + \sigma^{y y}}{2} = \frac{i e^2}{\hbar} \int \frac{d^2 k}{(2 \pi)^2} \frac{2 \omega \sqrt{k^2 + M^2}}{4 (k^2 + M^2)-\omega^2} \frac{k^2 + 2 M^2}{4(k^2 + M^2)^2} \, ,
\label{inthBN}
\eeq
where we utilized the formula \eqref{InterMetrCurv}. Before considering the insulating regime, $\omega < 2 M$, it is instructive to analyze the $M\to 0$ case first. The integrand is necessarily singular in this limit, and one needs to re-introduce the scattering rate to stabilize the conductivity $\omega \to \omega + i \varepsilon$. The value of the integral is then well-defined, and $\sigma^{xx}$ turns out to be purely real. Multiplying by four, which accounts for spin-valley degeneracy in graphene, we acquire the well-known result for the ac conductivity \cite{ac01, ac02}
\beq
\sigma^{xx} = \frac{\pi e^2}{2 h}\, .
\eeq
Remarkably, this contribution is both $\omega$- and $\varepsilon$-independent. 

In the opposite, insulating regime of vanishing frequency $\omega \to 0$ and a finite gap $M$, the system is characterized by a capacitive response $\sigma^{xx} = i \omega c_0$ with
\beq
\label{conek1App}
c_0 = \frac{e^2}{12 \pi |M|} \, .
\eeq
This simple expression has wide applicability: it describes the universal value of capacitance arising from a single Dirac cone with the mass $M$. For topologically trivial materials like hexagonal boron nitride, $c_0$ multiplied by the number of Dirac cones provides a good estimate for the capacitance (cf.  Fig.~2 in the main text). 

On the other hand, unlike the Hall conductivity $\sigma^{xy}$, $c_0$ is insensitive to the sign of $M$, and so, is unable to track the band inversions and subsequent changes in topology: it also fails to capture the capacitance deep in the topological phase. In order to understand how the intrinsic capacitance is influenced by topology, one can introduce a parabolic correction to the Dirac cone Hamiltonian \eqref{coneham}: $\hat H' = \hat H - \a k^2 \hat \s_z$. This model is characterized by a unit Chern number in the parameter region $\a M > 0$. The quantum metric traced over the spatial indices in this model takes the form:
\beq
g = \frac{k^2+2 M^2+2\a^2 k^4}{4 (k^2+M^2-2k^2 M \a + k^4 \a^2)^{2}}\, .
\eeq
The integral for conductivity can be performed analytically with the result for the capacitance
\beq
\label{conek2App}
c_0' = 
\begin{cases} 
        \frac{e^2}{12 \pi} \frac{1}{|M|  |1 - 4 M \alpha|}\, , & \a M < 0 \, , \\
      \frac{e^2}{12 \pi |M|} + \frac{|\alpha|}{6 \pi}\, , &  \a M > 0\, . \\
   \end{cases}
\eeq
Comparing this result with Eq.~\eqref{conek1App}, we find good agreement in the trivial region where $\a M<0$. 
However, in the topological region ($\a M<0$) with Chern number $\pm1$, the extended Dirac cone result Eq.~\eqref{conek2App} does not decay to zero as $|M| \to \infty$, instead saturating at $|\a|/6\pi$, in strong contrast to Eq.~\eqref{conek1App}. This reinforces the statement that the behavior of the intrinsic capacitance can serve as an indicator of the topology of the system.

Similar considerations apply to tight-binding models in a compact Brillouin zone. In order to illustrate this, we investigate a slight modification of the Haldane model
\beq
\label{HalHam}
\hat H_{\rm Haldane} = -\frac{2}{3}(f(k_x,k_y) \hat \s_+ + f^*(k_x,k_y) \hat \s_-)
+ M \hat \s_z
+ M_{\rm H} \tilde f(k_x,k_y) \hat \s_z
- \frac{3 \sqrt{3}}{2} M_{\rm H} \hat \s_z \, ,
\eeq

where 
\beq\bal
f(k_x,k_y) & = \sum_i e^{- i \bfa_i \cdot \bfk}\, , &\qquad& \bfa_1 = (0,-a),~ \bfa_{2,3} = (\pm \sqrt{3}a,a)/2 \, , \\
\tilde f(k_x,k_y) & = \sum_i \sin\( \bfb_i \cdot \bfk \)\, ,& \qquad &\bfb_1=(\sqrt{3}a,0)\, ,~\bfb_{2,3} = (-\sqrt{3}a,\pm 3 a)/2 \, ,
\eal\eeq
and we set for convenience $a=1$. The last term in Eq.~\eqref{HalHam} is introduced only for ease of notation: It shifts the topological region in parameter space, such that for $M_{\rm H}>0$, the parameter ranges $M<0$  and $M>3 \sqrt{3} M_{\rm H}$ correspond to trivial phases. In the range $0 < M < 3 \sqrt{3} M_{\rm H}$, the system acquires a finite Chern number $C=-1$. The resulting phase diagram, intrinsic capacitance and quantum metric are depicted in Fig.~2 of the main text.

\section{Dielectric constants of various materials}
\label{appendiel}

The dielectric constant in crystalline materials receives contributions both from ionic and electronic degrees of freedom \cite{BigDielBook}. Here, We are focusing on the latter, eliminating the contribution due to lattice dynamics by assuming a high enough value of the driving frequency $\omega$, while keeping it well below the gap $\Delta$.
In this regime, the ``slow'' lattice degrees of freedom remain inactive, whereas the electronic response remains entirely off-resonant, giving way to the quasi-static approximation assumed in Eq.~(6) of the main text.

\begin{table}[t!]
\centering

\begin{tabular}{ |p{1.8cm}|p{2.2cm}|p{1.7cm}|p{1.5cm}|p{1.5cm}|p{1.5cm}|p{1.2cm}|p{1.cm}|p{1.7cm}|p{1.7cm}|}
 \hline
 \multicolumn{10}{|c|}{Dielectric constants} \\
 \hline
 Material & Structure & Space group & Topology &$\e_{\inf}$, exp. & $\e_{\inf}$, th. & gap, $\rm eV$ & $a_z$, $\mathring{A}$& $2 \pi\langle\bar g\rangle_z$, exp. & $2 \pi \langle\bar g\rangle_z$, th. \\
\hline
$\rm Bi_2 Se_3$\textsuperscript{\cite{madelung, Kobayashi}}& \emph{rhombohedral}& $\rm{R\bar3m}$&TI  & 29  & 34.7 & 0.3 & 9.84& 2.87 & 3.45\\
$\rm Bi_2 Te_3$\textsuperscript{\cite{madelung, Kobayashi}}& \emph{rhombohedral}& $\rm{R\bar3m}$&TI   & 85  & 101.7 &0.13 & 10.5& 3.98 & 4.77 \\
$\rm Sb_2 Te_3$\textsuperscript{\cite{madelung}}& \emph{rhombohedral}& $\rm{R\bar3m}$&TI & 51  & ~---& 0.28 & 10.4& 5.06 & ~--- \\
\hline
$\rm MoS_2$\textsuperscript{\cite{munkbat, LiuTMD}} & \emph{van der Waals} & $ \rm P6_3/mmc$ & OAL\textsuperscript{\cite{TMDOAL}}  & 16.4    & $18.0$ & 1.82 & $3.13^\dagger$ & 3.05 & 3.36\\
$\rm WS_2$\textsuperscript{\cite{munkbat, LiuTMD}}& \emph{van der Waals}& $ \rm P6_3/mmc$ & OAL\textsuperscript{\cite{TMDOAL}}&   14.0  & $15.4$ &1.94 & $3.14^\dagger$  & 2.75 & 3.05\\
$\rm MoSe_2$\textsuperscript{\cite{munkbat, LiuTMD}}& \emph{van der Waals}& $ \rm P6_3/mmc$ & OAL\textsuperscript{\cite{TMDOAL}}   & 17.6 & $19.1$ &1.51 & $3.35^\dagger$& 2.92 & 3.18\\
$\rm WSe_2$\textsuperscript{\cite{munkbat, LiuTMD}}& \emph{van der Waals}& $ \rm P6_3/mmc$ & OAL\textsuperscript{\cite{TMDOAL}} &   15.8  & $16.2$ &1.59 & $3.36^\dagger$& 2.75 & 2.82\\
$\rm MoTe_2$\textsuperscript{\cite{munkbat, LiuTMD}}& \emph{van der Waals}& $ \rm P6_3/mmc$ & OAL\textsuperscript{\cite{TMDOAL}} & 22.6 & $22.1$& 1.03 & $3.62^\dagger$& 2.80 & 2.73\\
$\rm hBN$\textsuperscript{\cite{TMDdiel, hbnEXP}} &
\emph{van der Waals}& $ \rm P6_3/mmc$& trivial & 4.95 & ~--- & 5.97 & $2.51^\dagger$& 2.06 & ~---\\
\hline
$\rm CdTe$\textsuperscript{\cite{madelung}} & \emph{zincblende}& $\rm F\bar 4 3m$&  trivial & 7.1  & ~--- & 1.48 & 6.46& 2.03 & ~---\\
  $\rm GaAs$\textsuperscript{\cite{madelung}} & \emph{zincblende}& $\rm F\bar 4 3m$& trivial & 10.9  & ~--- & 1.42 & 5.65& 2.76 & ~---\\
 $\rm InP$\textsuperscript{\cite{madelung}}   &\emph{zincblende}& $\rm F\bar 4 3m$ & trivial & 10.9  & ~--- & 1.34 & 5.87& 2.70 & ~---\\
$\rm GaN$\textsuperscript{\cite{madelung}} & \emph{zincblende}& $\rm F\bar 4 3m$ & trivial &4.86  & ~--- & 3.17 & 4.53& 1.92 & ~---\\
$\rm ZnSe$\textsuperscript{\cite{madelung}} & \emph{zincblende}& $\rm F\bar 4 3m$& trivial & 5.7 & ~---& 2.82 & 5.67 &  2.61 & ~---\\
$\rm GaP$\textsuperscript{\cite{Wing2021}} & \emph{zincblende}& $\rm F\bar 4 3m$& trivial & 8.89 & ~---& 2.27 & 5.45& 3.39 & ~---\\
$\rm Si$\textsuperscript{\cite{Wing2021}} & \emph{diamond cubic}& $\rm Fd\bar 3 m$& OAL & 11.25  & ~---& 1.12 & 5.43& 2.16 & ~---\\
$\rm Ge$\textsuperscript{\cite{madelung}} & \emph{diamond cubic}& $\rm Fd\bar 3 m$& OAL & 16  & ~---& 0.81 & 5.66& 2.39 & ~---\\
$\rm C$\textsuperscript{\cite{Wing2021}} & \emph{diamond cubic}& $\rm Fd\bar 3 m$& OAL & 5.55 & ~---& 5.47 & 3.57& 3.09 & ~---\\
\hline
$\rm PbSe$\textsuperscript{\tiny \cite{madelung}} &
\emph{rocksalt}& $\rm Fm\bar 3m$& trivial  & 22.9 & ~---& 0.28 & 6.12& 1.30 & ~---\\
PbTe\textsuperscript{\tiny \cite{madelung, Baleva}}  &
\emph{rocksalt}& $\rm Fm\bar 3m$& trivial  & 34.8 & ~---& 0.31 & 6.46& 2.35 & ~---\\ 
SnTe\textsuperscript{\tiny \cite{madelung, Murase}}  &
\emph{rocksalt}& $\rm Fm\bar 3m$& TCI  & 55 & ~---& 0.36 & 6.33 & 4.27 & ~---\\ 
\hline
$\rm MgO$\textsuperscript{\cite{madelung}} & \emph{rocksalt}& $\rm Fm\bar 3m$& trivial &2.94 & ~---& 7.9 & 4.22& 2.25 & ~---\\
$\rm LiF$\textsuperscript{\cite{Villars2016}}&
\emph{rocksalt}& $\rm Fm\bar 3m$& trivial & 1.92 & ~--- & 14.2 & 4.03& 1.83 & ~---\\
$\rm NaCl$\textsuperscript{\cite{hbcp}} &
\emph{rocksalt}& $\rm Fm\bar 3m$& trivial & 2.38 & ~---& 8.97 & 5.64& 2.42 & ~---\\
 ${\rm SiO_2}$\textsuperscript{\cite{hbcp,Bonzi.Grad.2022}} & \emph{wurtzite} & $\rm P6_3mc $ & trivial & 2.19 & ~---& 8.50 & 5.40& 1.90 & ~---\\
 ${\rm ZnO}$\textsuperscript{\cite{hbcp,Bonzi.Grad.2022}} & \emph{wurtzite}& $\rm P6_3mc $ & trivial & 4.41 & ~---& 3.4 & 5.25& 2.11 & ~---\\
$\rm CaF_2$\textsuperscript{\cite{hbcp}} &
\emph{fluorite}& $\rm Fm\bar 3m$& trivial & 2.05 & ~---& 12.1 & 5.46& 2.41 & ~---\\

 \hline
\end{tabular}
\caption{
Electronic in-plane dielectric constants, values of the gap, and lattice constants $a_z$ for selected semiconductors. The theoretical values of $\e_\infty$ are obtained using the linear response expression \eqref{dielApp}. For transitional metal dichalcogenides, a 3-band tight-binding model \cite{LiuTMD} is used --- the comparison of the $2D$ ab initio with the $3D$ experimental values is justified by the weak dependence of the dielectric constant on the number of layers \cite{TMDdiel}. The theoretical values for $\rm Bi_2 Se_3$ and $\rm Bi_2 Te_3$ are obtained using a tight-binding simulation with Slater-Koster parameters given in \cite{Kobayashi}. ${}^\dagger$For van der Waals materials, $a_z$ refers to monolayer thickness, and the theoretical value of $\epsilon_\inf$ is computed for a monolayer.}
\label{Table}
\end{table}

In the following, we describe how the dielectric constant can be connected with the intrinsic capacitance $c$. To this end, note that Eq.~(6) in the main text, applied to the case of $3D$ materials, is closely related to the static dielectric susceptibility $\chi$ by a simple unit conversion factor of $\epsilon_0 \simeq 8.85 \cdot 10^{-12} \,\rm F/m$. To see this, one expresses the longitudinal conductivity as
\beq
\sigma^{xx} = \frac{j^x}{E_x} = \frac{I^x}{A} \frac{d}{V_x}  = \frac{d}{A}(i \omega C) = i \omega \frac{d C}{A} = i \omega \epsilon_0 \chi \, , 
\eeq
where we assumed a rectangular slab-shaped insulator with thickness $d$ and cross-section $A$, such that $C=\e\epsilon_0\chi A/d$. On the other hand, $\sigma^{xx}=i \omega c$, which implies $\chi=c/\epsilon_0$. The value of the dielectric constant $\e=1+\chi$ is therefore given by the following linear response expression (see also \cite{Grosso2000}):
\beq
\label{dielApp}
\e=1+ \frac{2 e^2}{\hbar \epsilon_0}   \sum_{m\neq n} \int_{\rm BZ}
f_n(1-f_m) \frac{ g^{xx}_{mn}}{\omega_{mn}} \, .
\eeq
Note that by considering only the horizontal component of the metric $g^{xx}$, we restrict ourselves to the in-plane permittivity usually termed $\e_{\parallel}$. Following \cite{TMDdiel}, for ab initio calculations of $\epsilon$ using 2D tight-binding models, the integral over the vertical dimension of the Brillouin zone in \eqref{dielApp} is replaced with the inverse monolayer thickness as
\begin{equation}
    \int dk_z \to \frac{2 \pi}{a_z}\, .
\end{equation}

The experimental and theoretical values for the materials presented in Fig.~4 of the main text are tabulated in Supplementary Table~\ref{Table}. We point out that the electronic component of the dielectric constant in the literature is commonly referred to as the \emph{optical dielectric constant} or the \emph{high-frequency dielectric constant}, and often denoted as $\epsilon_{\infty}$. 
For the sake of comparison with Fig.~4 of the main text, instead of using the rescaled susceptibility, we also plot the original product of the dielectric constant and gap ($\epsilon_{\inf}\Delta$) as a function of $\Delta$ in Supplementary Fig.~\ref{fig:epsnormal}. 
The difference is striking: 
while the materials belonging to different groups neatly align on $\left<\bar g\right>_z$ axis, it is much harder to explain the trends in $\epsilon_{\inf}\Delta$ since the dielectric constant depends on the unit cell volume and symmetry, in addition to the influence of the quantum geometry which we explored in the main text.

Lastly, we analyze how well the value of $\left<\bar g\right>_z$ obtained using the minimal gap $\Delta$
\begin{equation}
    \left<\bar g\right>_z=\frac{ \epsilon_0}{e^2} a_z \De (\epsilon-1) 
    \label{eq:estbd}
\end{equation}
compares with the actual values of $\left<g\right>_z$ obtained using the formula (14) in the main text using the tight-binding models \cite{LiuTMD, Kobayashi2014}. As one can infer from the first and third rows of \autoref{Table2}, the values of $\left<g\right>_z$ for topological insulators are much higher than the value of the lower bound, while for transitional metal dichalcogenides, the discrepancy is much milder. This difference can be explained by the fact that the topological insulators $\rm Bi_2 Se_3$ and $\rm Bi_2 Te_3$ are extremely dispersive, and the Brillouin zone integral in \eqref{dielApp} receives most of the contributions from larger values of $\Delta$. In order to obtain a better estimate of the quantum metric, some average value of the gap $\bar \Delta$ could be used. To this end, we write
\begin{equation}
    \left< g\right>_z \simeq \frac{ \epsilon_0}{e^2} a_z \bar \De (\epsilon-1) \, .
    \label{av_est}
\end{equation}
The most appropriate definition of $\bar \De$ is the gap at which the optical conductivity develops a maximum. This choice is motivated by the fact that the optical conductivity $\rm Re\, \sigma^{\mu \mu}$, a quantity proportional to the joint density of states, is related to the dielectric constant by a Kramers-Kronig relation
\begin{equation}
    \epsilon^{\mu \mu}=1 + \frac{2}{\pi \varepsilon_0} \int_0^\infty \frac{d \omega
}{\omega^2} {\rm Re}\, \s^{\mu \mu} \, .
\end{equation}
Because of the polynomial suppression of the high-frequency contribution in the integral above, good accuracy can be achieved by replacing $\hbar \omega$ under the integral with a gap $\bar \Delta$ corresponding to the first peak in the function $\rm Re \, \sigma^{\mu \mu}(\omega)$. Using this approximation with the expression for the real part of conductivity
\begin{equation}
{\rm Re}\, \s^{\mu \mu} = \f{\pi e^2} {\hbar} \int_{\rm BZ} \sum\limits_{n \neq m} \w_{mn}\, g^{\mu \mu}_{mn} \d(\w-\w_{mn})\, ,
\end{equation}
immediately reproduces the estimate \eqref{av_est}. For example, such frequency $\bar \Delta$ corresponds to the gap at the $M$-point in transitional metal dichalcogenides, where a logarithmic van Hove singularity is located. The optical conductivity curves of both $\rm Bi_2 Se_3$ and $\rm Bi_2 Te_3$ have a single pronounced peak at $\bar \Delta \simeq 2\, \rm eV$ for the first and $\bar \Delta \simeq 1\, \rm eV$ for the second material \cite{bisepeak, bitepeak}, and we use these values to estimate $\braket{g}_z$ with \eqref{av_est}. As one can infer from the first and the fifth rows in \autoref{Table2}, the values of the quantum metric obtained using \eqref{av_est} are in good agreement with the ab initio values (both are in bold). This shows that dielectric constants can be conveniently used to estimate the quantum metric once the average gap $\bar \Delta$ is identified from optical measurements or band structure arguments. Similar conclusion can be obtained for semiconductors, as shown in~\autoref{Table3}.

\begin{table}[h]

\begin{center}
\begin{tabular}{ |p{3.cm}|p{1.5cm}|p{1.5cm}|p{1.5cm}|p{1.5cm}|p{1.5cm}|p{1.5cm}|p{1.5cm}| } 
  \hline
  Material & $\rm MoS_2$ & $\rm WS_2$ & $\rm MoSe_2$ & $\rm WSe_2$ & $\rm MoTe_2$ & $\rm Bi_2 Se_3$ & $\rm Bi_2 Te_3$ \\ 
  \hline
  $2 \pi \langle g\rangle_z$, theory & \textbf{5.00} & \textbf{5.08} & \textbf{4.89} & \textbf{4.93} & \textbf{4.78} & \textbf{18.3} & \textbf{29.9}  \\ 
  \hline
  $\Delta$ (minimal gap) & 1.82 & 1.94 & 1.51 & 1.59 & 1.03 & 0.3 & 0.13  \\ 
  \hline
  $2 \pi \langle \bar g\rangle_z$, experiment & 3.05 & 2.75 & 2.92 & 2.75 & 2.80 & 2.87 & 3.98 \\ 
  \hline
  $\bar \Delta$ (average gap) & 2.72 & 3.44 & 2.29 & 2.89 & 1.74 & 2 & 1  \\ 
  \hline
  $2 \pi \langle g\rangle_z$, experiment & \textbf{4.55} & \textbf{4.88} & \textbf{4.42} & \textbf{4.99} & \textbf{4.72} & \textbf{19.1} & \textbf{30.6}   \\ 
  \hline
\end{tabular}
\caption{In the first row, the values of the quantum metric computed using the tight-binding models \cite{LiuTMD, Kobayashi}. The values of the minimal gap $\Delta$ and $\braket{\bar g}_z$ are copied from \autoref{Table}. The values of the energy gap at the $M$-point in the fourth row are found based on the tight-binding models for transitional metal dichalcogenides \cite{LiuTMD}. The estimates of the quantum metric obtained from experimental values of dielectric constants found in the fifth row are computed using \eqref{av_est}.}
\label{Table2}
\end{center}
\end{table}

\begin{table}[h]

\begin{center}
\begin{tabular}{ |p{3.cm}|p{1.5cm}|p{1.5cm}|p{1.5cm}|p{1.5cm}|p{1.5cm}| } 
  \hline
  Material & $\rm C$ & $\rm Si$ & $\rm Ge$ & $\rm GaAs$ & $\rm ZnSe$  \\ 
  \hline
  $2 \pi \langle g\rangle_z$, theory\textsuperscript{\cite{Sgiarovello2001}} & \textbf{9.19} & \textbf{12.2} & \textbf{14} & \textbf{11.5} & \textbf{8.28}   \\ 
  \hline
  $\Delta$ (minimal gap) & 5.47 & 1.12 & 0.81 & 1.42 & 2.82  \\ 
  \hline
  $2 \pi \langle \bar g\rangle_z$, experiment & 3.09 & 2.16 & 2.39 & 2.76 & 2.61  \\ 
  \hline
  $\bar \Delta$ (average gap)\textsuperscript{\cite{ravpenn}} & 13.6 & 4.8 & 4.3 & 5.2 & 7.1  \\ 
  \hline
  $2 \pi \langle g\rangle_z$, experiment & \textbf{7.67} & \textbf{9.28} & \textbf{12.7} & \textbf{10.1} & \textbf{6.57}  \\ 
  \hline
\end{tabular}
\caption{
Same comparison as in~\autoref{Table2}, but for a selection of semiconductors. The theory values are based on the first-principles calculation of Ref.~\cite{Sgiarovello2001}.}
\label{Table3}
\end{center}
\end{table}

\begin{figure*}
\includegraphics[width=.9\textwidth]{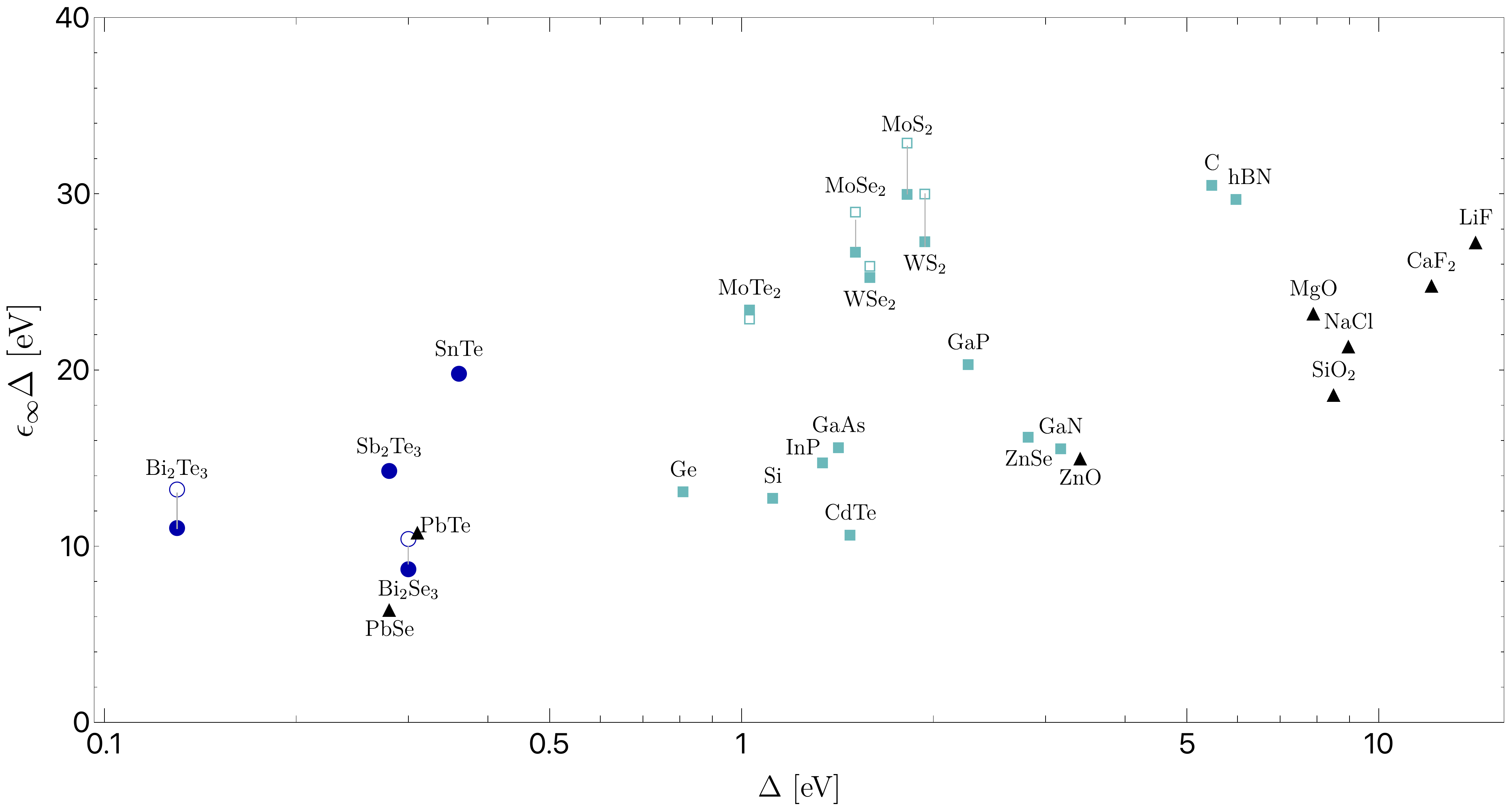}
\caption{Comparative figure demonstrating the conventional way of analyzing the dielectric constant as a function of the gap. Here we plot $\epsilon_{\inf}\Delta$, rather than $\left<\bar g\right>_z= a_z\Delta(\epsilon_\infty-1)$ (Fig.~4 of the main text). Note the absence of a trend: for instance, ionic trivial insulators denoted with black triangles span almost the entire range along the $y$-axis. In contrast, in Fig.~4 in the main text materials form tight groups.}
\label{fig:epsnormal}
\end{figure*}

\end{document}